\documentclass[twocolumn,english,superscriptaddress,amsmath,floatfix,longbibliography,floats,prl]{revtex4-1}
\usepackage[colorlinks=true,urlcolor=blue,citecolor=blue,linkcolor=blue]{hyperref} 
\usepackage[T1]{fontenc}
\usepackage[utf8]{inputenc}
\usepackage{amssymb}
\usepackage{graphicx}
\usepackage{amsmath,color}
\usepackage{mathrsfs}
\usepackage{float}
\usepackage{amsmath,bm}

\usepackage{indentfirst}
 \usepackage{txfonts}
\usepackage[normalem]{ulem}
\usepackage[table]{xcolor}
\usepackage{array,ragged2e}
\usepackage{grffile}
\usepackage{multirow}
\usepackage{algpseudocode}
\usepackage{epstopdf}

\usepackage[caption=false]{subfig}

\usepackage{graphicx,xcolor} 

\usepackage{comment}

\newcommand{\bra}[1]{\langle#1|}
\newcommand{\ket}[1]{|#1\rangle}
\newcommand{\norm}[1]{\langle#1|#1\rangle}
\newcommand{\ob}[2]{\langle#1|#2|#1\rangle}

\newcommand{\ele}[3]{\langle#1|#2|#3\rangle}

\newcommand{\blue}[1]{\textcolor{blue}{ #1}}

\begin{document}

\hyphenpenalty=5000
\tolerance=1000

\title{Accurate Simulation of the Hubbard Model with Finite Fermionic Projected Entangled Pair States}

\author{Wen-Yuan Liu}
\email{wyliu@zju.edu.cn}
 \affiliation{Division of Chemistry and Chemical Engineering, California Institute of Technology, Pasadena, California 91125, USA}
 \affiliation{Institute for Advanced Study in Physics, Zhejiang University, Hangzhou 310027, China}
\author{Huanchen Zhai}
 \affiliation{Division of Chemistry and Chemical Engineering, California Institute of Technology, Pasadena, California 91125, USA}
\author{Ruojing Peng}
 \affiliation{Division of Chemistry and Chemical Engineering, California Institute of Technology, Pasadena, California 91125, USA}
\author{Zheng-Cheng Gu}
 \affiliation{Department of Physics, The Chinese University of Hong Kong, Shatin, New Territories, Hong Kong, China}
\author{Garnet Kin-Lic Chan} 
\email{gkc1000@gmail.com}
 \affiliation{Division of Chemistry and Chemical Engineering, California Institute of Technology, Pasadena, California 91125, USA}

\begin{abstract}

We demonstrate the use of finite-size fermionic projected entangled pair states, in conjunction with variational Monte Carlo, to perform accurate simulations of the ground-state of the 2D Hubbard model. Using bond dimensions of up to $D=28$, we show that we can surpass state-of-the-art DMRG energies that use up to  $m=32000$ SU(2) multiplets on 8-leg ladders. We further apply our methodology to $10\times 16$, $12\times 16$ and $16 \times 16$ lattices at $1/8$ hole doping and observe the dimensional crossover between stripe orientations. Our work shows the power of finite-size fermionic tensor networks to resolve the physics of the 2D Hubbard model and related problems.

\end{abstract}
\maketitle

\date{\today}
\blue{\it Introduction.---}  The two-dimensional Hubbard model is one of the most intensively studied models in condensed matter physics~\cite{hubbard,Leblanc2015,zheng2017,wu2018pseudogap,jiang2019,qin2020,xiao2023}, and its ground state
is considered central to many fundamental many-body phenomena, including quantum magnetism, Mott insulators and high-temperature superconductivity~\cite{Dagotto1994,bulut2002,Scalapino2012,review1,review2}.
Computational methods have played a crucial part in revealing the physical nature of the Hubbard model~\cite{DMFT1,SCE1,CPT1,dmrg1992,CPAFQMC1,cQMC,DiaMC,DDMC,li2023tangent}, and recent advances provide evidence for the emergence of strange metallicity~\cite{huang2019strange}, the pseudogap~\cite{pseudogap}, and the dome-like structure of superconductivity~\cite{xu2024}. In parallel, the Hubbard model has also been realized in recent cold atom  quantum simulations~\cite{greif2016site,cheuk2016observation,mazurenko2017cold,sompet2022realizing,bourgund2025formation} and further suggested as an effective model for nickelate superconductivity~\cite{nickelate2019,nickelate2020}. However, due to its intrinsic complexity, achieving accurate numerical simulations away from half-filling remains challenging.
 
 The density matrix renormalization group (DMRG) method is recognized as a reliable method for the Hubbard model~\cite{dmrg1992,gong2021robust,li2023tangent,xu2024,lu2024,qu2024phase}, but accurate simulations are limited to pseudo-one-dimensional instances. Projected entangled pair states (PEPS)~\cite{PEPS2004,verstraete2008,fPEPS2010} are a higher-dimensional generalization of the DMRG ansatz (the matrix product state (MPS)~\cite{mps1992}), and have evolved into an advanced numerical tool. In recent years, progress has been made with calculations using infinite-sized PEPS (iPEPS), which directly access the thermodynamic limit as a function of the unit cell in the ansatz~\cite{jordan2008classical,jiang2008,roman2009,phien2015,corboz2016,vanderstraeten2016,cheng2018,liao2019,corboz2023efficient}.  Concurrently, finite-size PEPS methods, which do not assume a unit cell and complement iPEPS, have also been advanced~\cite{lubasch2014,liu2017,dong2019,haghshenas2019conversion,zaletel2020,liu2021,schneider2021,gray2024,liu2024tensor}, and have contributed to resolving the nature of the ground-state of 2D frustrated spin systems~\cite{liu2022gapless,liu2022emergence,liu2024emergent,liu2024j1j2j3,liu2024quantum}. 

\begin{figure}[tbp]
 \centering
 \includegraphics[width=3.4in]{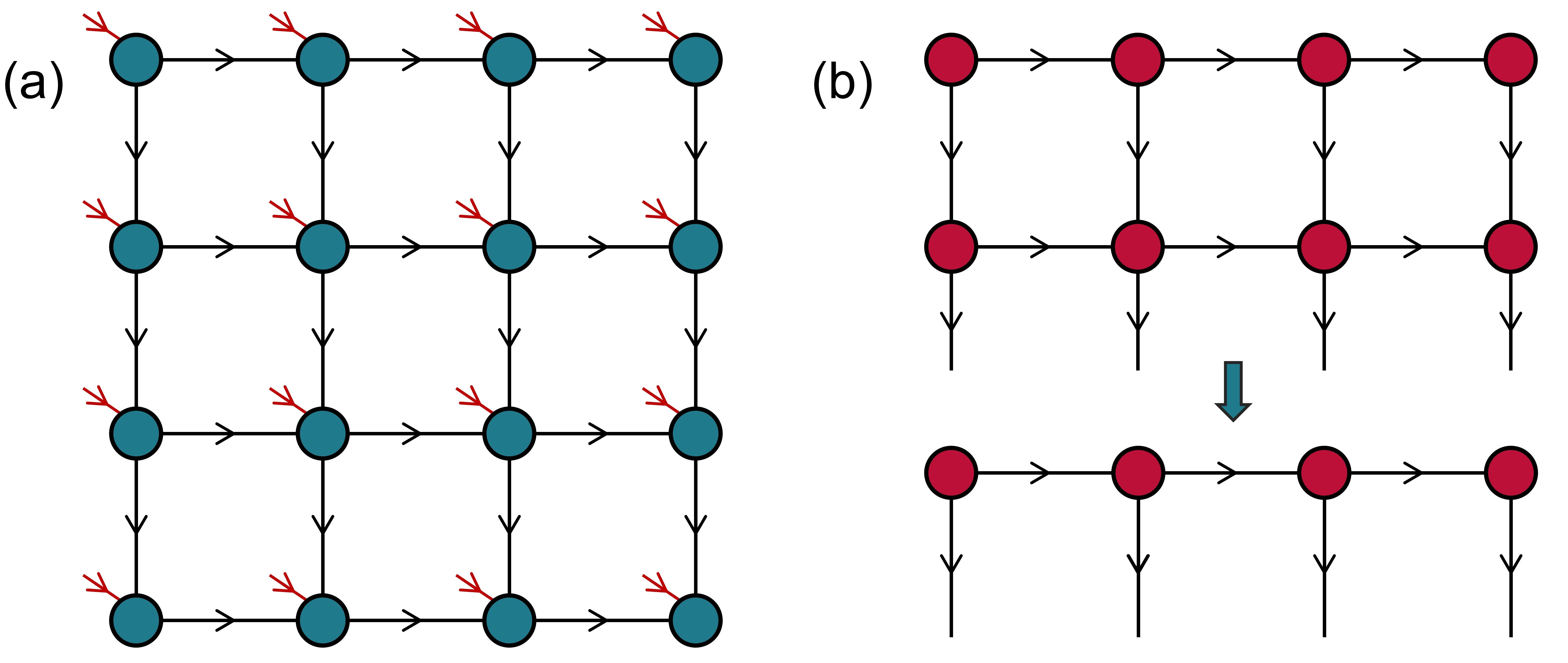}
 \caption{ (a) Graphical representation of fPEPS $\ket{\Psi}$ on a $4\times 4$ square lattice. Red legs denote the physical degrees of freedom which are of dimension 4 for the Hubbard model, and black legs are of dimension $D$ and control the expressive power of the fPEPS. Arrows between two connected tensors denote the relative order in the fermionic contraction. (b) For a given configuration $\ket{\bf k}$, the standard boundary-MPS method is used to compute its amplitude $\Psi({\bf k})$ by contracting the corresponding single-layer tensor network.}
 \label{fig:fPEPS}
 \end{figure}

Despite the promise of the PEPS ansatz for fermionic lattice ground states~\cite{fPEPS2010,corboz2010simulation,corboz2011,corboz2014competing,dong2020stable}, and some studies that have employed infinite PEPS for the 2D Hubbard model
~\cite{corboz2016improved,corboz2023}, its effectiveness for finite 2D Hubbard lattices has not yet been demonstrated, as the few works in the literature show a large gap between the accuracy of PEPS and that of DMRG on relevant lattice sizes~\cite{scheb2023}. Yet finite PEPS simulations remain desirable for several reasons. They alleviate the bias of the iPEPS unit cell, which is relevant for the long wavelength, inhomogeneous phases of the 2D Hubbard model, making finite PEPS a natural choice for studying many systems with broken translation-invariance~\cite{liu2024j1j2j3}, including disordered systems and optical lattice cold atoms. Finite PEPS also carry the advantage that they allow for a direct comparison and cross-check with methods such as DMRG and Quantum Monte Carlo which also work with finite lattices, which is crucial to resolve challenging regimes~\cite{liu2022gapless,liu2024quantum}. In addition, they can be related to efforts in neural networks and  machine learning, which also work in the finite setting~\cite{cheng2021,vieijra2022generative,liu2024tensor}. 
The challenges for accurate finite PEPS simulation are: (i) the accessible PEPS bond dimension $D$, which controls the representational ability, and (ii) the optimization strategy for large lattices, given the tensors are no longer constrained by a unit cell. Finite PEPS have traditionally been studied in the context of deterministic approximate contraction algorithms, where the largest  bond dimension $D$  previously described on square lattices is $D=8$~\cite{lee2023evaluating,scheb2023}. In the 2D Hubbard model, ground state energy accuracies of (better than) 1\% are required to distinguish between competing phases, and this bond dimension is insufficient.

In this work, we demonstrate a finite PEPS approach that realizes an accurate simulation of the 2D Hubbard model ground state. Specifically, using a fermionic PEPS (fPEPS) ansatz
within a variational Monte Carlo algorithm, we demonstrate simulations with a bond dimension up to $D=28$, surpassing the accuracy of the largest practical DMRG calculations on the 8-leg ladders to which DMRG is usually applied in the Hubbard model, and extending the simulations to sizes inaccessible to DMRG, including up to $16 \times 16$ lattices, where we observe the dimensional crossover of the stripe pattern. These results thus substantially expand the capability of finite PEPS and establish it as a powerful numerical tool for the 2D Hubbard model and other challenging problems in strongly correlated electron systems.

\blue{\it General setup and optimization.---}  The Hubbard model Hamiltonian on a  square lattice reads:
\begin{equation}
\label{eq:hubbard}
 H=-t\sum_{\langle ij \rangle,\sigma}(c^{\dagger}_{i,\sigma}c_{j,\sigma}+ h.c.)+ U\sum_{i}n_{i\uparrow}n_{i\downarrow}-\sum_{i}\mu_i (n_{i\uparrow}+n_{i\downarrow})-\sum_{i}h_iS^z_i,
\end{equation}
where $\sigma=\uparrow$ $(\downarrow)$ denotes spin up (down), $c^{\dagger}_{i,\sigma}$ ($c_{i,\sigma}$) is the electron creation (annihilation) operator on site $i$, $n_{i\uparrow}$ ($n_{i\downarrow}$) is the electron number operator for spin up (down),  $S^z_i$ is the spin-$z$ operator, and $\langle ij \rangle$ indicates nearest neighbors. We set $t=1$ and $U=8$, relevant to cuprates, and consider variational states where the total electron number $N_e$ is a good quantum number. We set the chemical potential $\mu_i=0$ ($h_i=0$) except in the case of charge (magnetic) pinning fields, where $\mu_i$ ($h_i$) is set to a non-zero value on certain sites in the initial optimization to establish a given order. 

The fPEPS ansatz on the square lattice~\cite{fPEPS2010}, is illustrated in Fig.~\ref{fig:fPEPS}(a). By defining appropriate fermionic contraction rules, fPEPS defines the wave function amplitude as a contraction of tensors associated with each lattice site similar to a conventional bosonic tensor network~\cite{PEPS2004}, but with additional local tensor operations that account for fermionic anti-commutation~\cite{barthel2009,corboz2010simulation,grassmann2010,pizon2010,dong2019}. The current work uses contraction rules following the Grassmann tensor network formalism~\cite{grassmann2010,grassmann2013}.
To represent a wavefunction with a fixed electron number $N_e$, we impose U(1) charge symmetry on the local tensors~\cite{singh2011tensor,bauer2011}. 

In conventional PEPS calculations, physical quantities are computed deterministically by contracting a double-layer tensor network comprised of the bra and ket.  The computational cost scaling is $O(D^6\chi_d^2)+O(D^4\chi_d^3)$ on square lattices, where $\chi_d$ is an approximation bond dimension for double-layer contraction, typically chosen as $\chi_d= O(D^2)$, which leads to a scaling of $O(D^{10})$~\cite{verstraete2008,lubasch2014}. 
Here we employ the variational Monte Carlo (VMC) sampling technique with fPEPS, which replaces the summation over the physical degrees of freedom by an importance sampling~\cite{sandvik2007,schuch2008,wang2011,wouters2014,liu2017,dong2019,liu2021} of the wavefunction amplitudes $\Psi({\bf k})=\langle {\bf k} \ket{\Psi}$, $\ket{{\bf k}}=\ket{k_1k_2 \cdots k_N}$ for $N$ sites.
In this approach, only a single-layer tensor network needs to be contracted to compute  $\Psi({\bf k})$, which we do using the boundary-MPS contraction method [Fig.~\ref{fig:fPEPS}(b)]. The leading cost scaling of fPEPS-VMC is then $O(M D^4\chi_s^{2})+O(M D^4\chi_s^{3})$, where the latter arises from boundary-MPS contraction using SVD compression, but has a small prefactor and is not dominant in our calculations. $\chi_s$ is an approximation bond dimension for single-layer contraction, typically chosen as $\chi_s= O(D)$ (see SM~\cite{SM}), which eventually leads to a scaling of $O(M D^6)$. 
$M$ is the number of MC sweeps~\cite{liu2021}, typically on the order of $10000$ for all sizes studied, and each sweep involves $O(N)$ samples~\cite{liu2021}.
All physical quantities can then be evaluated by sampling, including the energy gradients with respect to the tensor elements needed for gradient optimization of fPEPS.

When optimizing the fPEPS wave function, we first use the simple update (SU)  method, which has a cost scaling of $O(D^5)$~\cite{jiang2008,wang2011}. During the SU optimization, we optionally add temporary pinning fields to certain sites {($\mu_i$ or $h_i$ in Eq.~\ref{eq:hubbard})}. After converging the SU optimization, we further use gradient-based optimization (GO) methods to improve the accuracy in a subset of simulations~\cite{liu2017,liu2021}. After optimization, the physical quantities are measured by MC sampling. 
Out of the various GO methods, we have found the stochastic reconfiguration (SR) method~\cite{SR1998,Neuscamman2012,directSam2021} to work the best (see SM~\cite{SM}), thus the GO results presented are from SR optimization.
Key factors enabling the accurate simulations in this work are the efficiency of SU for the orders of the 2D Hubbard model, the accessibility of large bond dimension fPEPS through MC sampling, and the use of pinning fields to establish different local orders. In the following, we will illustrate the impact of these factors.

\begin{figure}[tbp]
 \centering
 \includegraphics[width=0.5\textwidth]{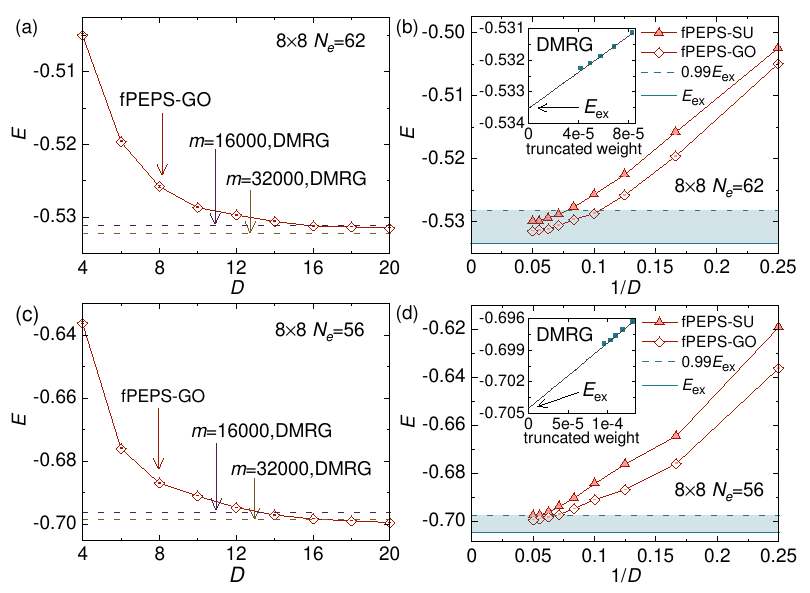}
 \caption{ The energies (per site) of the $8\times 8$ Hubbard model at $U=8$ with electron numbers $N_e=62$  (a-b)  and  $N_e=56$ (c-d). (a) and (c) present the energies from GO (gradient-based optimization) with respect to  fPEPS bond dimension $D$. DMRG results with  $m=16000$ and $32000$ SU(2) multiplets are shown for comparison. (b) and (d) present the $1/D$ dependence of the fPEPS energies from SU (simple update, filled triangles) and GO (unfilled diamonds). The border lines of the shaded region correspond to the extrapolated DMRG energies $E_{\rm ex}$  (solid) and  $0.99E_{\rm ex}$ (dashed).  Insets of (b) and (d) show the extrapolation of the DMRG energies w.r.t. the truncated weights from bond dimension $m=16000$, 20000, 24000, 28000. The energies of $m=32000$ are also shown. The fPEPS energy error bars are smaller than the symbols.} 
 \label{fig:8x8}
 \end{figure}

\begin{figure*}[tbp]
 \centering
 \includegraphics[width=0.99\textwidth]{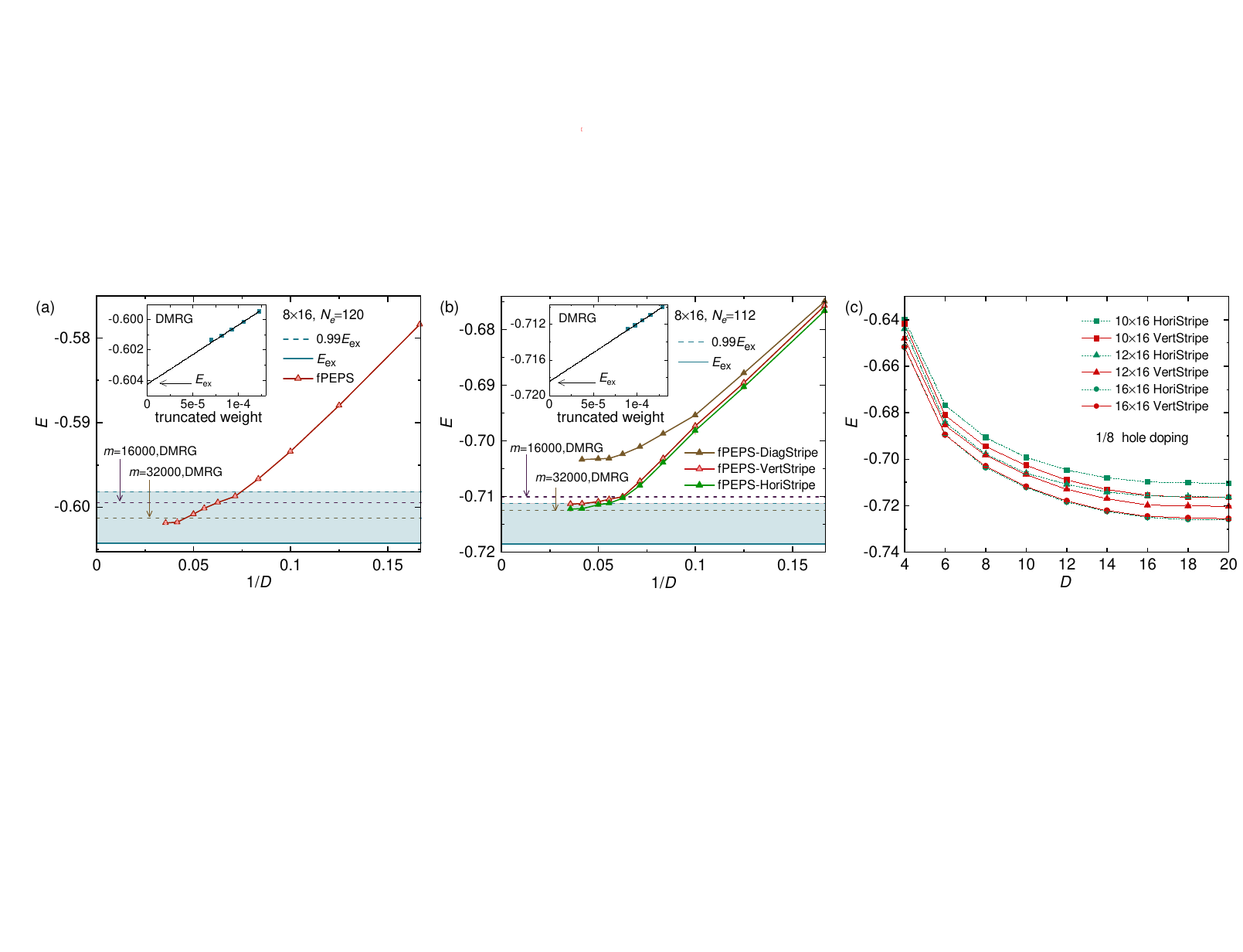}
 \caption{ Energies of the $8\times 16$ Hubbard model at $U=8$ with (a) $N_e=120$ and (b) $N_e=112$, using fPEPS with a bond dimension up to $D=28$.
 In (b), fPEPS energies with diagonal (brown), vertical (violet) and horizontal (red) stripe states are shown for comparison.    DMRG energies with $m=16000$ and $32000$ SU(2) multiplets are also shown. Insets of (a) and (b) show the extrapolations of DMRG energies w.r.t. the truncated weights using bond dimension $m=16000$, 20000, 24000, 28000 (i.e. not including $m=32000$). (c) Energies of the horizontal and vertical stripes for the $10\times 16$, $12\times 16$, $16\times 16$ lattices at $U=8$, with electron numbers $N_e=140$, $168$, $224$, respectively, i.e., $n_h=1/8$. The fPEPS energy error bars are smaller than the symbols.} 
 \label{fig:Energy_all}
 \end{figure*}

We use open boundary conditions (OBC) along both the $x$ and $y$ directions, which allows for a direct comparison between fPEPS and DMRG. In the DMRG calculations U(1)$\times$SU(2) symmetry for the charge and spin degrees of freedom was used, and the  bond dimension $m$ in the DMRG calculations denotes the number of SU(2) multiplets.

\blue{\it Benchmarks on $L\times L$ systems.---} We first start with benchmarks on small square lattices. Using converged DMRG energies as the reference, for $4\times4$ at half-filling $n_h=1$ and at the hole doping concentration $n_h=0.125$, the relative energy errors of
fPEPS with $D=16$ are $1.6\times 10^{-5}$ and $4.4\times 10^{-4}$, respectively; for $6\times6$ at $n_h=1/18$ and $n_h=1/9$, the relative energy errors of fPEPS $D=20$ are $ 1.6 \times 10^{-3}$ and $4.4\times 10^{-3}$, respectively. See SM~\cite{SM}. All of these are in good agreement with the DMRG results, and could be further improved if desired, as the calculations are not expensive.

\begin{figure*}[tbp]
\centering
 \includegraphics[width=0.99\textwidth]{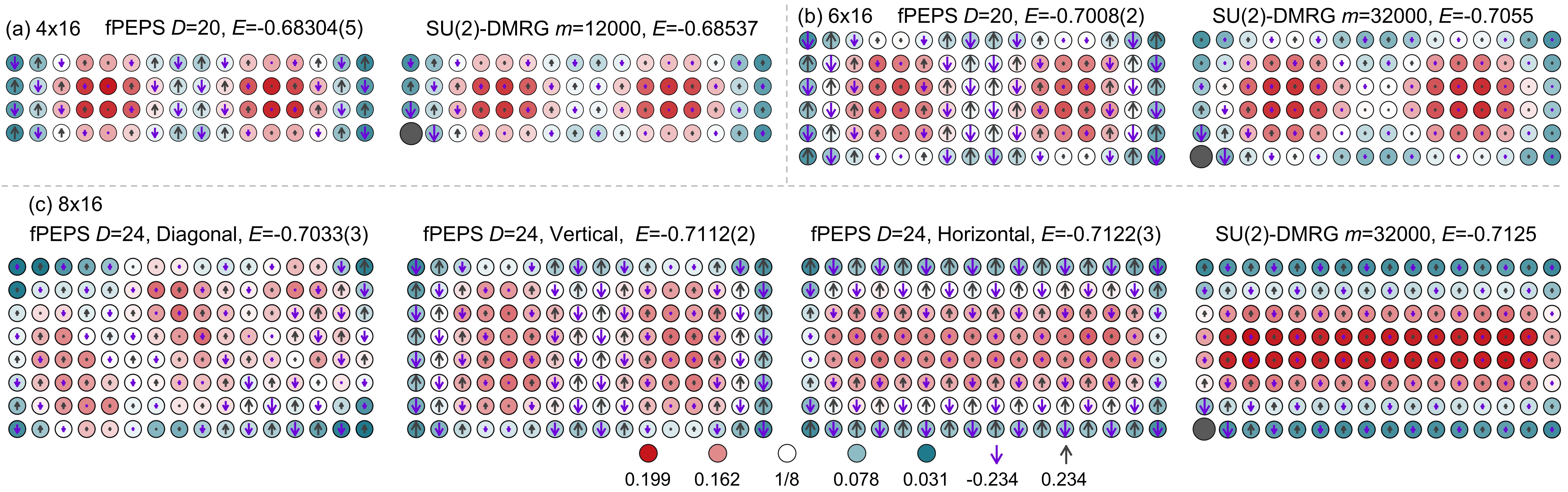}
\caption{Hole density and spin moment (local spin $z$-component for fPEPS and spin correlations for DMRG) distributions on OBC systems including (a) $4\times 16$,  (b) $6\times 16$,  and (c) $8\times 16$  lattices at $U=8$ with hole doped concentration $n_h=1/8$, i.e. with electron number $N_e=56$, 84 and 112, respectively. 
For DMRG results, local spin $z$-component moments are zero due to SU(2) symmetry, and the presented arrows denote the spin correlation values $|\frac{1}{3}\langle {\bf S}_{\rm ref}\cdot {\bf S}_{i,j}\rangle|^{1/2}$ using the reference site $(0,0)$ (filled black circles), and each arrow direction (up or down) shows the sign of $\langle {\bf S}_{\rm ref}\cdot {\bf S}_{i,j}\rangle$. The background hole density $n_h=1/8$ is subtracted to emphasize the fluctuation of the hole density around the background value. The magnitudes of the spin moments are represented by the sizes of the arrows. In (c) for $8\times 16$, three types of stripes are found in the fPEPS optimization, including the diagonal, the vertical (along $y-$ direction) and the horizontal (along $x-$ direction).  The horizontal stripe has the lowest energy, within statistical error of the lowest DMRG energy. Sampling errors of spin moments and hole densities from fPEPS are respectively around 1\% and 6\% of the corresponding mean values. } 
\label{fig:patternSmall}
\end{figure*}

\begin{figure}[tbp]
\centering
 \includegraphics[width=0.5\textwidth]{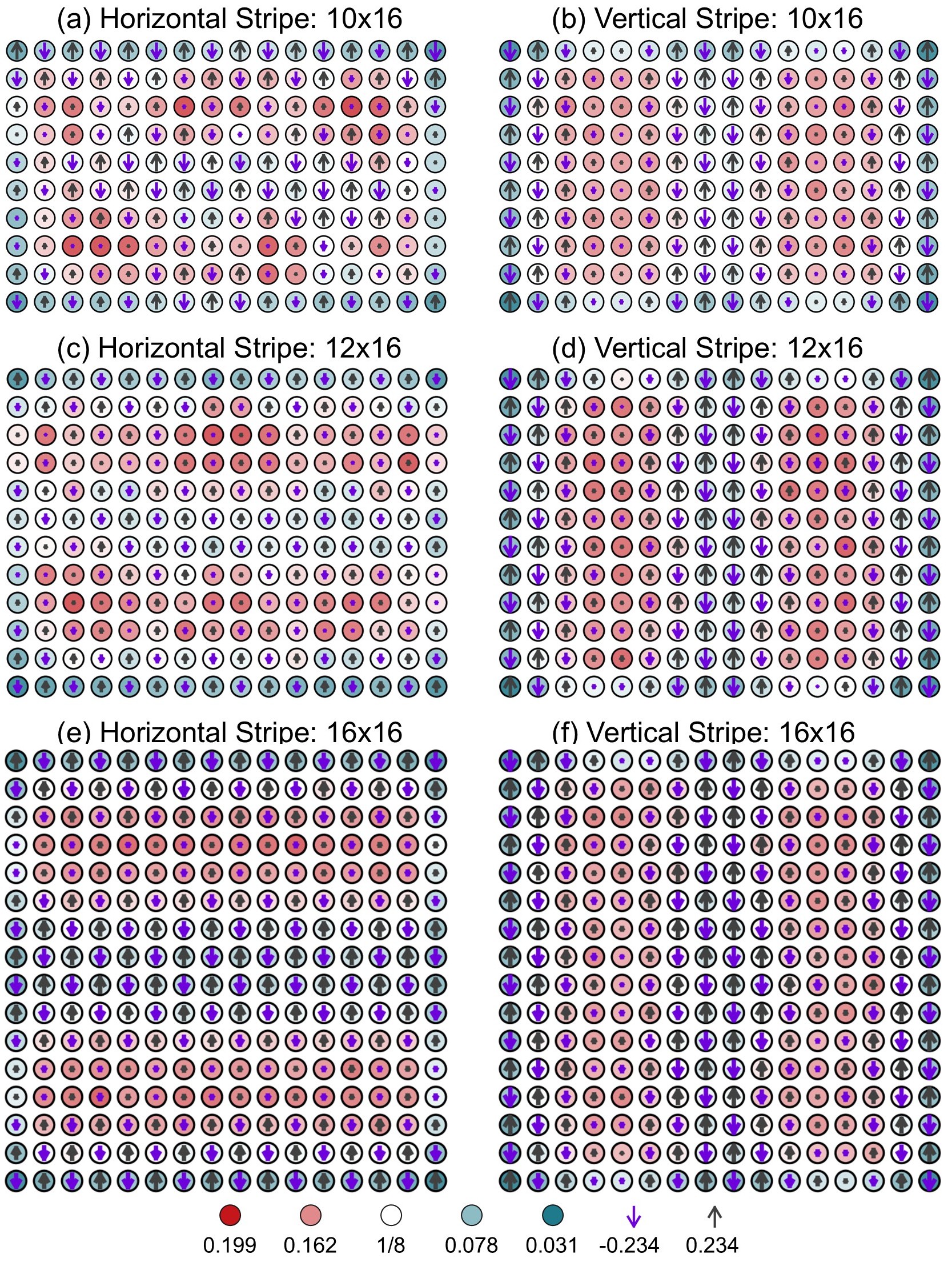}
\caption{Horizontal and vertical stripe patterns for different lattice sizes for $U=8$ and hole doping $n_h=1/8$, using fPEPS with $D=20$. For the $10\times 16$ and $12\times 16$ lattices, the vertical stripe has lower energies than the horizontal ones, and for $16\times 16$, horizontal and vertical stripes are degenerate; see Fig.~\ref{fig:Energy_all}(c). Sampling errors of spin moments and hole densities from fPEPS are respectively around 1\% and 6\% of the corresponding mean values. } 
\label{fig:patternLarge}
\end{figure}

In Fig.~\ref{fig:8x8}, we present the fPEPS energies for the $8 \times 8$ lattice, obtained using both GO and SU optimization methods with varying $D$, compared to large DMRG calculations. As shown in Figs.~\ref{fig:8x8}(a), \ref{fig:8x8}(b), for $N_e=62$, the energy with $D=20$ is lower than the variational DMRG result for $m=16000$, with a relative error of $4 \times 10^{-3}$ compared to the extrapolated DMRG ($m=\infty$) estimate. For $N_e=56$ (Figs.~\ref{fig:8x8}(c), \ref{fig:8x8}(d)) the energy with $D=20$ using SU optimization is $-0.6975$, comparable to the DMRG energy with $m=24000$ SU(2) multiplets. With further gradient optimization, the fPEPS energy improves to $-0.69928(3)$, lower than the energy of the largest variational DMRG calculation, $-0.69840$ with $m=32000$, with a relative error of $7 \times 10^{-3}$  compared to the extrapolated DMRG estimate, $-0.70449(129)$. The uncertainty in the extrapolated DMRG energy is reported as one fifth of the extrapolation distance (the difference between the extrapolated and $m = 28000$ energies for the current case)\cite{olivares2015ab}. 

In practice, we found that for the $8 \times 8$ system with $N_e = 56$, simple SU optimization often gets trapped in a local minimum characterized by a diagonal stripe, and further performing GO does not escape this local minimum.
To overcome this, we used charge pinning fields with $\mu=0.5$. For example, to stabilize a vertical stripe, we applied this field to the two middle rows of the system, starting from a random state with $D=8$. Once the SU optimization for $D=8$ was converged (see SM~\cite{SM}), we turned off the pinning field and continued with the SU optimization, keeping the fPEPS bond dimension at $D=24$ until convergence. Other smaller-$D$ fPEPS states were then obtained through a reverse process, sequentially decreasing
$D$ from 24 to smaller values, and ensuring the convergence of the SU optimization for each given $D$. The use of temporary pinning fields is similar to that in other DMRG optimizations~\cite{zheng2017}.
In this way we were able to investigate the competition of different ordered states.

\blue{\it Simple update vs Gradient optimization.---}  Very interestingly, by comparing the energy errors of the GO and SU methods,  we find that the energy error of SU for a given $D$ is only roughly twice as large as that of GO, see SM for details~\cite{SM}.
 Meanwhile, we also observe that both SU and GO yield nearly identical patterns for the local orders for large $D$ (see SM~\cite{SM}), with only slight quantitative differences, as seen in 
 Fig.~\ref{fig:patternSmall}(b) for the charge and spin moments in the $6\times 16$ lattice with fPEPS $D=20$, compared to those from DMRG. 
Because of this, for larger lattices we have only used SU optimization, which is much cheaper than GO, allowing for a large bond dimension $D$ (here, up to $D=28$). This is advantageous in the combination with MC sampling, which enables the physical quantities of the large-$D$ fPEPS from the SU to be efficiently evaluated.

\blue{\it Results on $8\times 16$ systems.---} $8$-leg ladders are commonly regarded as the widest systems accessible to DMRG simulations for the 2D Hubbard model. Indeed, for the $8\times 16$ ladder, even when using $m=32000$ SU(2) multiplets, the DMRG calculations show 
large truncated weights, namely $7.0\times 10^{-5}$ and $8.9\times 10^{-5}$ respectively for $N_e=120$ ($n_h=1/16$)  and  $N_e=112$ ($n_h=1/8$), as illustrated in the insets of Fig.~\ref{fig:Energy_all}.

Figures~\ref{fig:Energy_all}(a) and (b) present the fPEPS energies from SU optimization up to $D=28$.  At $n_h=1/16$, the energy decreases noticeably with increasing $D$ until $D=24$ and improves only slightly from $D=24$ to $28$. Note that the fPEPS $D=16$ energy is close to the DMRG energy with $m=16000$, and the $D=24$ energy is  $-0.60176(25)$, below the variational DMRG energy $-0.60137$ with $m=32000$. 
The lowest fPEPS energy with $D=28$ is $-0.60188(19)$ which is within  a relative error of $4\times 10^{-3}$ of the extrapolated DMRG estimate of $-0.60426(64)$.

At doping $n_h=1/8$ 
which is of great interest~\cite{zheng2017}, the DMRG  $m=32000$ calculations finds a horizontal stripe, seen in  the hole density and spin correlation functions shown in Fig.~\ref{fig:patternSmall}(c). For fPEPS, we first conducted the SU optimization without pinning fields, resulting in a diagonal stripe. This diagonal stripe is at a  high energy, as shown in Fig.~\ref{fig:Energy_all}(b). We then performed SU by adding temporary charge pinning fields to the central two rows to induce a similar horizontal stripe pattern to the DMRG ground-state, and the energy of this horizontal stripe with $D=24$ is essentially the same (within statistical error) as that from DMRG using $m=32000$. For the vertical and horizontal stripes, it is notable that the energy decreased very quickly from $D=4$ to $D=16$, changed slowly from $D=16$ to $D=24$, and showed minimal improvement from  $D=24$ to $D=28$.

Since we found the vertical stripe phase to be the ground-state on the $4\times 16$ and $6\times 16$ ladders instead of the horizontal one seen above,
we also imposed a temporary magnetic pinning field ($|h_i|=0.5$)  in the SU process to converge to a vertical striped state. From Fig.~\ref{fig:Energy_all} (b) we can see that at the same $D$ value the vertical stripe is slightly higher in energy than the horizontal stripe, indicating that the horizontal stripe indeed is favored on the $8\times 16$ lattice. 
However, the energy difference between horizontal and vertical stripes is small ($\sim 0.001$) and it is thus of interest to investigate their competition on larger lattices.

\blue{\it fPEPS results on larger sizes.}  The above detailed comparisons with DMRG explicitly demonstrate the reliability of our fPEPS results. Now we consider larger sizes including $L_y\times 16$ with $L_y=10$, $12$, and $16$, which are beyond the scope of accurate DMRG simulations. 
 The energies of the stable horizontal and vertical stripe patterns for each fPEPS $D$ are shown in Fig.~\ref{fig:Energy_all}(c), and the spin and hole density patterns for $D=20$ are presented in Fig.~\ref{fig:patternLarge}. Note that, in the case of the horizontal stripes, the pinning fields were applied to the middle two rows ($y=L_y/2$ and $L_y/2+1$), but SU optimization relaxed the distribution to approximately $y=L_y/4$ and $y=3L_y/4$, resulting in the horizontal stripe pattern in Fig.~\ref{fig:patternLarge}.

 In contrast to  $8\times 16$, we find for $10\times 16$ and $12\times 16$, that the vertical stripes are favoured over the horizontal ones, as seen in Fig.~\ref{fig:Energy_all}(c). For $16\times 16$, horizontal and vertical stripes have essentially the same energy; this is reasonable since the $x-$ and $y-$ directions are equivalent, and this serves as another check on the correctness of our results. The overall findings support a horizontal/vertical stripe phase with wave length $\lambda =8$, consistent with previous studies~\cite{zheng2017}, with a dimensional crossover between the two as a function of system width.

\blue{\it Conclusions.---} In summary, we have demonstrated the power of finite-size PEPS for the 2D Hubbard model, establishing that it achieves state-of-the-art accuracy by direct comparison with DMRG energies on narrow systems, and demonstrating the ability to reach large lattice sizes, such as the $16 \times 16$ lattice. Our calculations at 1/8 doping support the stability of the wavelength $8$ stripe order seen in other studies, but allow us to further demonstrate a dimensional crossover between horizontal and vertical stripes as a function of system width. Our findings open the door towards resolving longstanding questions about fermionic ground-states via the use of finite PEPS, including questions of superconductivity in the 2D Hubbard model and related systems, while also providing a powerful classical approach for benchmarking and complementing quantum simulators.

\blue{\it Acknowledgment.---} The DMRG calculations in this work were performed using \textsc{block2}\cite{zhai2021low, zhai2023block2, block2}, and the scripts can be found in \url{https://github.com/hczhai/2d-hubbard-dmrg-2024}. The computations presented in this work were conducted at the Resnick High Performance Computing Center, a facility supported by the Resnick Sustainability Institute at the California Institute of Technology. This work was primarily supported by the U.S. Department of
Energy, Office of Science, National Quantum Information Science Research Centers, Quantum Systems Accelerator. Additional support for HZ (DMRG calculations) was provided by US Airforce Office of Scientitic Research under award AFOSR-FA9550-18-1-0095. GKC is a Simons Investigator in Physics. ZCG is supported by funding from Hong Kong's Research Grants Council (CRF C7012-21GF and RGC Research Fellow Scheme 2023/24, No. RFS2324-4S02).
Wen-Yuan Liu also acknowledges additional support from a start-up grant from Zhejiang University for the final part of this work.


\appendix
\setcounter{equation}{0}
\newpage

\renewcommand{\thesection}{S-\arabic{section}} \renewcommand{\theequation}{S%
\arabic{equation}} \setcounter{equation}{0} \renewcommand{\thefigure}{S%
\arabic{figure}} \setcounter{figure}{0}

\centerline{\textbf{Supplemental Material}}

\section{I. Grassmann tensor networks}

There are different equivalent formalisms for fermionic tensor network computations~\cite{barthel2009,corboz2010simulation,grassmann2010,fPEPS2010,pizon2010,dong2019}. Here we use the Grassmann representation~\cite{grassmann2010,grassmann2013}, which is equivalent to the $\mathbb{Z}_2$-graded tensor representation~\cite{mortier2024fermionic}, in which all fermionic tensor operations correspond to local tensor operations. For a discussion of the relationship between this ``local'' formulation and a ``global'' ordering formulation, see Ref.~\cite{gao2024fermionic}. For clarity and completeness, here we briefly review fermionic tensor networks in the Grassmann representation.

\subsection{A. fermionic PEPS}

We begin with the definition of the fermionic PEPS on a square lattice following Ref.~\cite{fPEPS2010}. We first consider the product state composed of a series of entangled fermion pairs on all links:
\begin{equation}
\ket{\Psi_f}=\prod_{\langle i,j \rangle} (1+a_{\rm X_i}^{\dagger}a_{\rm X_j}^{\dagger})\ket{0}=\prod_{\langle i,j \rangle}{\bf G}_{ij}\ket{0},
\label{eq:directPro}
\end{equation}
where $\langle i,j \rangle$ denotes the link $X$ (nearest neighbor) between sites $i$ and $j$ , and $a_{\rm X_i}^{\dagger}$ and  $a_{\rm X_j}^{\dagger}$ are  fermionic creation operators on sites $i$ and $j$ at the ends of the link $X$. We call the fermions created by $a_{\rm X_i}^{\dagger}$ and $a_{\rm X_j}^{\dagger}$, virtual fermions, by convention. 

Now we map the virtual fermionic space on site $i$ to the physical fermionic space via the projector:
\begin{equation}
{\bf P}_i=\sum_{m_i}\sum^{1}_{lurd=0}T^{[m_i]}_{lurd}c^{\dagger m_i}a_{\rm L_i}^{l}a_{\rm U_i}^{u}a_{\rm R_i}^{r}a_{\rm D_i}^{d},
\label{eq:projector}
\end{equation}
where $c^{\dagger}$ is the physical fermionic creation operator at site $i$, and $a_{\rm X_i}$ $(\rm X=L,U,R,D)$ is the fermionic annihilation operator on site $i$ connecting its neighbouring sites in the left, up, right or down directions. For simplicity, we consider the physical and virtual fermionic degrees of freedom to both be of dimension $2$, i.e., $m$ and $x$ $(x=l,u,r,d)$ are 0 or 1. To ensure that the parity of the PEPS is well defined, we can assume all elements $T^{[m_i]}_{lurd}$ are zero if ($l+u+r+d+m_i$) is odd. 
Then, the fermionic PEPS is expressed as:
\begin{equation}
\ket{\Psi}=\Big\langle \prod_i {\bf P}_i\prod_{\langle i,j \rangle}{\bf G}_{ij} \Big\rangle_0 \ket{0}
\label{eq:fPEPS}
\end{equation}
Here $\langle\cdots \rangle_0$ denotes the expectation of the virtual fermions over the vacuum (i.e. integrating out the virtual fermionic degrees of freedom). 
The fPEPS bond dimension here corresponds to $D=2$.

For general fPEPS with a bond dimension $D$, by introducing multiple fermionic modes, we can  write down the formal generalization as:
\begin{align}
&\ket{\Psi_f}=\prod_{\langle i,j \rangle} \Big(\sum^{D-1}_{n=0}a_{{\rm X_i},n}^{\dagger p(n)}a_{{\rm X_j},n}^{\dagger p(n)}\Big)\ket{0}=\prod_{\langle i,j \rangle}{\bf G}_{ij}\ket{0}, \\
& {\bf P}_i=\sum_{m_i}\sum_{lurd=0}^{D-1} T^{[m_i]}_{lurd}c^{\dagger m_i}a_{{\rm L_i},l}^{p(l)}a_{{\rm U_i},u}^{p(u)}a_{{\rm R_i},r}^{p(r)}a_{{\rm D_i},d}^{p(d)},
\end{align}
Here $a^{\dagger}_{{\rm X_i},n}$ is the fermionic operator to create a fermionic mode, and $p(n)$ is either 0 or 1 following the parity of this fermionic mode. 

Specifically, imagine there are $z$ fermions with $2^z$ associated fermion modes,  and $a_{{\rm X_i},n}^{\dagger}\ket{0} = \ket{n_1n_2\cdots n_z}$  ($n_k=0$ or 1) with a parity $p(n)={\rm mod}(\sum_{k=1}^{z}{n_k},2)$. The bond dimension $D$ represents the $D$ fermionic modes corresponding to the $2^z$ different occupancies, and the subscript $n$ in $a_{{\rm X_i},n}^{\dagger}$ distinguishes the different modes. Since only the even or odd parity matters for the fermionic signs,  $a_{{\rm X_i},n}^{\dagger}$ works basically as a conventional fermionic operator, and the fPEPS still has the form Eq.~(\ref{eq:fPEPS}). As mentioned above $\ket{\Psi}$ is required to have a well defined parity, say even parity, thus we  assume all elements $T^{[m_i]}_{lurd}$ are zero if $[p(m_i)+p(l)+p(u)+p(r)+p(d)]$ is odd.

When contracting fermionic tensor networks, we need to be careful with the order of the fermionic operators, and any change in order requires following the anticommutation relations. Below we adopt the Grassmann formalism in which the possible signs caused by anticommutators can be computed locally~\cite{grassmann2010,grassmann2013}.

\subsection{B. Grassmann Tensor Representations}

\subsubsection{1. Grassmann algebra}
In the standard Grassmann algebra, for the Grassman variable $\eta_i$ and its conjugate $\bar\eta_i$, we have
\begin{align}
    &  \eta_i\eta_j=-\eta_j\eta_i,\quad  \bar\eta_i\bar\eta_j=-\bar\eta_j\bar\eta_i, \quad \eta_i\bar\eta_j=-\bar\eta_j\eta_i,
    \label{eq:grassmann_a} \\
    &\int{\rm d}\eta_i\eta_j= \int{\rm d}\bar\eta_i\bar\eta_j=\delta_{ij}.
        \label{eq:grassmann_b}
\end{align}

Considering two Grassmann functions ${\bf A(\xi,\bar \eta)}=A_{00}+A_{10}\xi+A_{01}\bar\eta+A_{11}\xi\bar\eta$ and ${\bf B(\eta, \beta)}=B_{00}+B_{10}\eta+B_{01}\beta+B_{11}\eta\beta$, the scalar product obtained by integrating out the variable $\eta$ and its conjugate $\bar\eta$ is defined with a Grassmann metric ${\rm e}^{-\eta\bar\eta}$~\cite{negele1998},
\begin{equation}
\begin{split}
&\int_{\eta}{\bf A(\xi,\bar \eta)}{\bf B(\eta, \beta)} =\int{\rm d}\eta{\rm d}\bar\eta {\rm e}^{-\eta\bar\eta} {\bf A(\xi,\bar \eta)}{\bf B(\eta, \beta)}\\
&=\int{\rm d}\eta{\rm d}\bar\eta (1-\eta\bar\eta)(A_{00}+A_{10}\xi+A_{01}\bar\eta+A_{11}\xi\bar\eta) \\
&\times(B_{00}+B_{10}\eta+B_{01}\beta+B_{11}\eta\beta)\\
&=A_{00}B_{00}+A_{01}B_{10}+(A_{10}B_{00}+A_{11}B_{10})\xi \\
& +(A_{00}B_{01}+A_{01}B_{11})\beta+ (A_{10}B_{01}+A_{11}B_{11})\xi\beta
\label{eq:grassmann_product_standard}
\end{split}
\end{equation}

In the above calculation, the existence of the symbol ${\rm d}\eta{\rm d}\bar\eta$ and the metric ${\rm e}^{-\eta\bar\eta}$, makes the Grassmann integral computations in Eq.~(\ref{eq:grassmann_product_standard})  somewhat tedious. We find the action of ``$\int{\rm d}\eta{\rm d}\bar\eta {\rm e}^{-\eta\bar\eta}$ '' can be conveniently realized using a  simple rule defined as
\begin{equation}
     \int\bar\eta_i^{p(i)}\eta_j^{p(j)}=\delta_{ij}\delta_{p(i),p(j)}, ~ \int\eta_j^{p(j)}\bar\eta_i^{p(i)}=(-1)^{p(i)}\delta_{ij}\delta_{p(i),p(j)}.
    \label{eq:newGrelation}
\end{equation}
The subscript $i$ in $\bar\eta_i$ specifies the species of Grassmann variable, and $p(i)=0$ or $1$ is its parity, denoting the absence or presence of the variable.

With this, we reconsider Eq.~(\ref{eq:grassmann_product_standard}). We first write the Grassmann functions  ${\bf A(\xi,\bar \eta)}$ and ${\bf B(\eta,\beta)}$ in a Grassmann tensor form, i.e.,  ${\bf A(\xi,\bar \eta)}=\sum_{l_1l_2}A_{l_1l_2}\xi^{p(l_1)}\bar\eta^{p(l_2)}$ and ${\bf B(\eta,\beta)}=\sum_{l_3l_4}B_{l_3l_4}\eta^{p(l_3)}\beta^{p(l_4)}$, where $p(l_i)$ is 0 (1) when the Grassmann variable is absent (present). Using Eq.~(\ref{eq:newGrelation}), then the integral over $\eta$ and its conjugate $\bar\eta$ can be directly read out
\begin{equation}
\begin{split}
&\int_{\eta}{\bf A(\xi,\bar \eta)}{\bf B(\eta, \beta)}\\
&=\int\sum_{l_1l_2l_3l_4}\Big(A_{l_1l_2}\xi^{p(l_1)}\bar\eta^{p(l_2)}\Big)\Big( B_{l_3l_4}\eta^{p(l_3)}\beta^{p(l_4)}\Big) \\
&=\sum_{l_1l_2l_3l_4}A_{l_1l_2}B_{l_3l_4}\xi^{p(l_1)}\Big(\int\bar\eta^{p(l_2)} \eta^{p(l_3)}\Big)\beta^{p(l_4)}\\
&=\sum_{l_1l_2l_3l_4}A_{l_1l_2}B_{l_3l_4}\xi^{p(l_1)}\delta_{l_2l_3}\beta^{p(l_4)}\\
&=\sum_{l_1l_2l_4}A_{l_1l_2}B_{l_2l_4}\xi^{p(l_1)}\beta^{p(l_4)},
\label{eq:grassmann_product_new}
\end{split}
\end{equation}
which is identical to Eq.~(\ref{eq:grassmann_product_standard}), showing the validity of Eq.~(\ref{eq:newGrelation}).

Therefore, with the help of Eq.~(\ref{eq:newGrelation}), we can replace the Grassmann algebra of Eqs.~(\ref{eq:grassmann_a}-\ref{eq:grassmann_b}) by the following relations, which are simpler for our fermionic computations:
\begin{align}
   &  \eta_i\eta_j=-\eta_j\eta_i,\quad  \bar\eta_i\bar\eta_j=-\bar\eta_j\bar\eta_i, \quad \eta_i\bar\eta_j=-\bar\eta_j\eta_i,
       \label{eq:Newgrammann_a} \\
    &\int\bar\eta_i^{p(i)}\eta_j^{p(j)}=\delta_{ij}\delta_{p(i),p(j)}, ~ \int\eta_j^{p(j)}\bar\eta_i^{p(i)}=(-1)^{p(i)}\delta_{ij}\delta_{p(i),p(j)}.
      \label{eq:Newgrammann_b} 
\end{align}
Here $p(i)$ is either 0 or 1, denoting the absence or presence of the Grassmann variable. Intuitively, the Grassmann relations realize the action of the fermionic operators $a^{\dagger}_i$ and $a_i$ through the Grassmann variables $\eta_i$ and its conjugate $\bar\eta_i$. The integral in Eq.~(\ref{eq:Newgrammann_b}) is similar to the expectation value of $ a_i a^{\dagger}_j $ over the vacuum, and mimics the sum over virtual fermionic degrees of freedom in fermionic tensor networks. Using the basic relations Eqs.~(\ref{eq:Newgrammann_a}-\ref{eq:Newgrammann_b}), the following rules can be derived directly.

\subsubsection{2. Grassmann tensor operations}

A general rank-$r$ Grassmann tensor containing $m$ $\eta$-type variables and $(r-m)$ $\bar{\eta}$-type variables can be expressed as:
\begin{equation}
{\bf T}=\sum_{l_1l_2..l_r}T_{l_1l_2..l_r}\eta_{{\rm X_1},l_1}^{p(l_1)}\cdots\eta_{{\rm X_m},l_m}^{p(l_m)}\bar{\eta}_{{\rm X_{m+1}},l_{m+1}}^{p(l_{m+1})}\cdots\bar{\eta}_{{\rm X_r},l_r}^{p(l_r)}.
\end{equation} 
 $\rm X_i$ labels the Grassmann numbers on different tensor legs, and on a given leg $\rm X_i$, $l_i$ is used to distinguish different Grassmann variables on this leg, and $p(l_i)$ is the corresponding parity, either 0 or 1. $T_{l_1l_2..l_r}$ is a number, i.e. the tensor element. When working with the Grassmann tensor ${\bf T}$, we not only need to deal with the elements $T_{l_1l_2..l_r}$, but also consider the order of the Grassmann variables.  In the following we will omit the label of leg $X_i$ in the expression of ${\bf T}$ for convenience:
\begin{equation}
{\bf T}=\sum_{l_1l_2..l_r}T_{l_1l_2..l_r}\eta_{l_1}^{p(l_1)}\cdots\eta_{l_m}^{p(l_m)}\bar{\eta}_{l_{m+1}}^{p(l_{m+1})}\cdots\bar{\eta}_{l_r}^{p(l_r)} .
\label{eq:Tbasic}
\end{equation}

{\it Tensor parity}.  Generally, we can choose Grassmann  tensors to have a definitive parity $p({\bf T})$, that is  all tensor elements $T_{l_1l_2..l_r}$ satisfy:
\begin{equation}
   T_{l_1l_2..l_r}=0, ~\quad {\rm if}~\quad  {\rm mod} \Big(  p(l_1)+p(l_2)+\cdots +p(l_r), 2 \Big)=1-p({\bf T}).
   \nonumber
\end{equation}
If $p({\bf T})=0$ (1),  we say the tensor has an even (odd) parity.

Note for the odd-parity tensor $\bf T$ with $p({\bf T})=1$, it can always be converted into an even-parity tensor $\tilde{\bf T}$ with  $p(\tilde{\bf T})=0$ by introducing an extra tensor leg. Specifically, for an odd-parity tensor  $\bf T$,
\begin{equation}
   T_{l_1l_2..l_r}=0, ~\quad {\rm if}~\quad  {\rm mod}\Big( p(l_1)+p(l_2)+\cdots +p(l_r), 2\Big)=0.
      \nonumber
\end{equation}
After adding an extra tensor leg $k$, 
\begin{equation}
   \tilde{T}_{l_1l_2..l_rk}=0, ~\quad {\rm if}~\quad  {\rm mod}\Big( p(l_1)+p(l_2)+\cdots +p(l_r)+k, 2\Big)=1,
      \nonumber
\end{equation}
where the leg $k$ is dimension-1 with an odd parity. Then the tensor  $\tilde{\bf T}$ comprised of $\tilde{T}_{l_1l_2..l_rk}$ has an even parity. Such a parity changing operation is very convenient for practical fermionic tensor network computations~\cite{corboz2010simulation}.

{\it Permutation.} 
For a Grassmann tensor, if we exchange the positions of any two Grassmann variables  $\eta_{l_t}$ and $\eta_{l_{t+1}}$, the resulting Grassmann tensor is 
\begin{align}     
&{\bf T'}=\sum_{l_1..l_r}T_{l_1..l_r}\eta_{l_1}^{p(l_1)}\cdots \eta_{l_t}^{p(l_t)}\eta_{l_{t+1}}^{p(l_{t+1})}\cdots \bar\eta_{l_r}^{p(l_r)} \nonumber\\
 &=\sum_{l_1..l_r}(-1)^{[p(l_t)*p(l_{t+1})]}T_{l_1..l_r}\eta_{l_1}^{p(l_1)}\cdots \eta_{l_{t+1}}^{p(l_{t+1})}\eta_{l_t}^{p(l_t)}\cdots \bar\eta_{l_r}^{p(l_r)} .
 \end{align}
 This is basically the result of the anti-commutation relation, as shown in Eq.~(\ref{eq:Newgrammann_a}).  From the permutation rule, the following Hermitian conjugate and decomposition/contraction rules can be directly derived.

{\it Hermitian conjugate.} The Hermitian conjugate of Grassmann tensor $\bf T$ in Eq.~(\ref{eq:Tbasic}) is 
\begin{align}
   & {\bf T}^{\dagger}=\sum_{l_1..l_r}\tilde{T}_{l_1l_2..l_r}\eta_{l_r}^{p(l_r)}\cdots \eta_{l_{m+1}}^{p(l_{m+1})} \bar \eta_{l_m}^{p(l_m)} \cdots \bar \eta_{l_1}^{p(l_1)},  
   \label{eq:conjugate_a}\\
   & \tilde{T}_{l_1l_2..l_r} =(-1)^{\sum_{i=m+1}^rp(l_i)}T^{*}_{l_1l_2..l_r}. 
 \label{eq:conjugate_b}  
\end{align}
 Note that $\bf T$ has a definite parity and thus $ p({\bf T})=\sum_{i=1}^r p(l_i) \mod 2$. 

We can motivate Eqs.~(\ref{eq:conjugate_a}-\ref{eq:conjugate_b}). We know ${\bf T}^{\dagger}$ should have the form Eq.~(\ref{eq:conjugate_a}), and thus we only need to derive  Eq.~(\ref{eq:conjugate_b}). We consider the scalar product  $\int_{\rm all}{\bf T^{\dagger}T}$ (i.e. the norm of the Grassmann tensor), which is expected to be a real number $\sum_{l_1l_2..l_r}|T_{l_1l_2..l_r}|^2$ by integrating out all indices. On the other hand, according to Eq.~(\ref{eq:Newgrammann_b}), we have $\int\bar{\eta}^{p(l_k)}_{l_k} \eta^{p(l_k)}_{l_k}=1$ and $\int \eta^{p(l_k)}_{l_k} \bar{\eta}^{p(l_k)}_{l_k} =(-1)^{p(l_k)}$. It is easy to verify that $\int_{\rm all}{\bf T^{\dagger}T}=\sum_{l_1l_2..l_r}T_{l_1l_2..l_r} \tilde{T}_{l_1l_2..l_r} (-1)^{\sum_{i=m+1}^{r}p(l_i)}$, and thus we obtain Eq.~(\ref{eq:conjugate_a}).

Using  Eqs.~(\ref{eq:conjugate_a}-\ref{eq:conjugate_b}), we derive the identity $\int_{\rm all}{\bf TT^{\dagger}}=\sum_{l_1l_2..l_r}|T_{l_1l_2..l_r}|^2 (-1)^{\sum_{i=1}^{r}p(l_i)}=(-1)^{p(\bf T)} \int_{\rm all}{\bf T^{\dagger}T}$, where the total parity $p({\bf T})$ determines the sign. This sign reversal arises because exchanging two odd-parity tensors ($p({\bf T})=1$) introduces a factor of $-1$, while even-parity tensors ($p({\bf T})=0$) leave the integral invariant, thereby confirming the consistency of the conjugate relations.  Additionally, the double conjugate satisfies $({\bf T}^\dagger)^\dagger=(-1)^{p({\bf T})}{\bf T}$. These observations  suggest that handling even-parity Grassmann tensors is more convenient than their odd-parity counterparts.

\begin{figure}[tbp]
 \centering
 \includegraphics[width=3.4in]{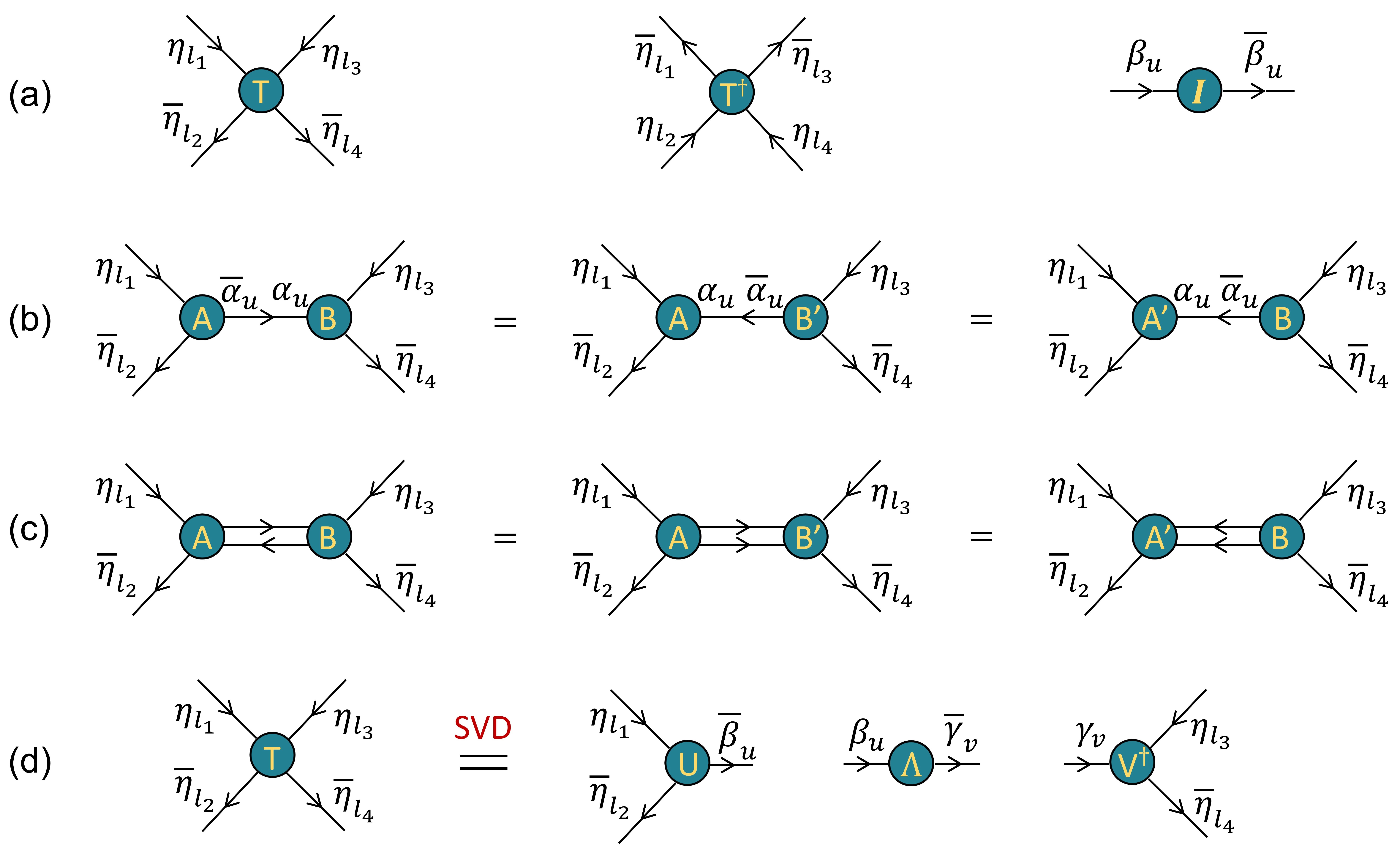}
 \caption{ Graphical notation for Grassmann tensors. Incoming and outgoing arrows correspond to Grassmann variables $\alpha$ and $\bar{\alpha}$, respectively. (a) Grassmann tensor ${\bf T} $ and its Hermitian conjugate. (b) Grassmann tensor contractions for one common index. (c) Grassmann tensor contractions for two common indices with different arrow directions. (d) SVD for a Grassmann tensor.}
 \label{fig:GrassmannTensor}
 \end{figure}

As a convenient graphical notation, we  assign an arrow to each tensor leg,  shown in Fig.~\ref{fig:GrassmannTensor}(a). The tensor legs with  incoming arrows correspond to  ${\alpha}$ (creation operator $c^{\dagger}$), and outgoing arrows correspond to $\bar{\alpha}$ (annihilation fermionc operator $c$). For its Hermitian conjugate ${\bf T}^{\dagger}$, we need to reverse the arrow direction for each leg, and modify the tensor elements by multiplying by the phase factor $(-1)^{p(l_i)}$ for each outgoing leg of ${\bf T}$ according to Eq.~(\ref{eq:conjugate_b}).  

{\it Decomposition and contraction.} Here we first look at the rule of decomposition. Defining  ${\bf T}= \sum_{l_1\dots l_r}T_{l_1 \dots l_r} \eta_{l_1}^{p(l_1)}\cdots \eta_{l_r}^{p(l_r)}$, then  $T_{l_1 \dots l_r}$ can be decomposed as $T_{l_1\dots l_r}=\sum_{uv}A_{l_1\cdots l_t, u}\delta_{uv}B_{v,l_{t+1}\cdots l_r}$.  According to Eq.~(\ref{eq:Newgrammann_b}), using $\int{\bar \alpha^{p(u)}}_u\alpha^{p(v)}_v=\delta_{uv}$ (here $u=0,1,\cdots, D-1$ distinguishes the $D$ Grassmann variables and $p(u)$ is the corresponding parity and thus $\delta_{uv}=\delta_{p(u),p(v)}$), we  have ${\bf T}=\int_{\alpha} {\bf AB}$ (integrating out the $\alpha$ Grassmann variable), where 

\begin{equation}
 \begin{split}
     &{\bf A}=\sum_{l_1..l_t,u}A_{l_1\cdots l_t, u}\eta_{l_1}^{p(l_1)}\cdots \eta_{l_t}^{p(l_t)}\bar \alpha^{p(u)}_u ,
\label{eq:GA} \\
&{\bf B}=\sum_{l_{t+1}..l_r,u}B_{u,l_{t+1}\cdots l_r}\alpha^{p(u)}_u \eta_{l_{t+1}}^{p(l_{t+1})}\cdots\eta_{l_r}^{p(l_r)}.
 \end{split}   
\end{equation}

Similarly, we can  also use $\int \alpha^{p(v)}_v {\bar \alpha^{p(u)}}_u=(-1)^{p(u)}\delta_{uv}$ and absorb the phase factor $(-1)^{p(u)}$ into $\bf B$, and then we have ${\bf T}=\int_{\alpha} {\bf AB'}$ where

\begin{equation}
 \begin{split}
&{\bf A}=\sum_{l_1..l_t,u}A_{l_1\cdots l_t, u}\eta_{l_1}^{p(l_1)}\cdots \eta_{l_t}^{p(l_t)} \alpha^{p(u)}_u ,
\label{eq:GA_1} \\
&{\bf B'}=\sum_{l_{t+1}..l_r,u}(-1)^{p(u)}B_{u,l_{t+1}\cdots l_r}\bar\alpha^{p(u)}_u \eta_{l_{t+1}}^{p(l_{t+1})}\cdots\eta_{l_r}^{p(l_r)}.
 \end{split}   
\end{equation}

Of course, we can also absorb the phase factor $(-1)^{p(u)}$ into $A$ rather than $B$, and then we have ${\bf T}=\int_{\alpha} {\bf A'B}$ where

\begin{equation}
 \begin{split}
&{\bf A'}=\sum_{l_1..l_t,u}(-1)^{p(u)}A_{l_1\cdots l_t, u}\eta_{l_1}^{p(l_1)}\cdots \eta_{l_t}^{p(l_t)} \alpha^{p(u)}_u ,
\label{eq:GA_2} \\
&{\bf B}=\sum_{l_{t+1}..l_r,u}B_{u,l_{t+1}\cdots l_r}\bar\alpha^{p(u)}_u \eta_{l_{t+1}}^{p(l_{t+1})}\cdots\eta_{l_r}^{p(l_r)}.
 \end{split}   
\end{equation}

 Contraction can be viewed as the reverse of decomposition. Taking the Grassmann tensor ${\bf T}=\sum_{l_1l_2l_3l_4}T_{l_1l_2l_3l_4}\eta^{p(l_1)}_{l_1}\bar\eta^{p(l_2)}_{l_2}\eta^{p(l_3)}_{l_3}\bar\eta^{p(l_4)}_{l_4}$ as an example, the above three different decomposition forms in Eq.~(\ref{eq:GA}), Eq.~(\ref{eq:GA_1}) and Eq.~(\ref{eq:GA_2}) resepectively correspond to the contraction of two tensors, shown in Fig.~\ref{fig:GrassmannTensor}(b). In practice, the tensor legs to be contracted may have different arrow directions, shown in Fig.~\ref{fig:GrassmannTensor}(c). In this situation, we can arrange the arrows to have the same direction by modifying the tensor elements, according to Eq.~(\ref{eq:GA_1}) or (\ref{eq:GA_2}). 

The singular value decomposition (SVD) is another basic operation in tensor network computations. As an application, here we look into how to perform SVD.  According to conventional SVD, we have $T_{l_1l_2l_3l_4}=\sum_{uv}U_{l_1l_2,u}\Lambda_{uv}(V^{\dagger})_{v,l_3l_4}$. Then the Grassmann SVD for ${\bf T}=\int_{\beta\gamma} {\bf U}  \Lambda {\bf V}^{\dagger}$ can be directly written down:

\begin{equation}
 \begin{split}
    & {\bf U}=\sum_{l_1l_2u}U_{l_1l_2,u}\eta^{p(l_1)}_{l_1}\bar\eta^{p(l_2)}_{l_2}\bar\beta_u^{p(u)} \\
    & \Lambda=\sum_{uv}\Lambda_{uv}\beta_u^{p(u)} \bar\gamma_v^{p(v)} \\
    & {\bf V^{\dagger}}=\sum_{vl_3l_4}(V^{\dagger})_{v,l_3l_4}\gamma_v^{p(v)}\eta^{p(l_3)}_{l_3}\bar\eta^{p(l_4)}_{l_4}
 \end{split}   
\end{equation}

We  find $\bf U$ and $\bf V^{\dagger}$ naturally satisfy the unitary property. Taking $\bf U$ as an example, according to Eq.~(\ref{eq:conjugate_a}) and the unitary properties of $U_{l_1l_2,u}$, it is easy to verify that 
\begin{align}
    \int_{\eta_{l_1}\eta_{l_2}}{\bf U^{\dagger} U}=(-1)^{p(\bf U)}\int_{\eta_{l_1}\eta_{l_2}}{\bf UU^{\dagger}}=\sum_{u}\beta_{u}^{p(u)}\bar\beta_{u}^{p(u)}, 
\end{align}
corresponding to the $D\times D$ {\it identity} Grassmann tensor where $D$ is the number of Grassman variables ($u=0,1,\cdots, D-1$), similar to Eq.~(\ref{eq:Newgrammann_b}). The QR (LQ) decomposition can be performed in the same way.

Therefore, in practice, for Grassmann tensors, we only need do tensor contractions or decompositions on the tensor elements $T_{l_1\cdots l_r}$, and simultaneously consider possible phase factors arising from the anti-commutation relations. All these operations are local, and thus Grassmann tensors can be very conveniently dealt with like bosonic tensors.

\subsection{C. Grassmann representation for fPEPS}

Replacing fermionic operators  by Grassmann variables, we can express Eq.~(\ref{eq:directPro}) and Eq.~(\ref{eq:projector}) with their Grassmann forms: 
\begin{align}
&{\bf G}_{ij}=\sum^{D-1}_{s=0}  \eta_{{\rm X_i},s}^{p(s)} \eta_{{\rm X_j},s}^{p(s)}\ket{0},
\label{eq:GeneralGform} \\
&{\bf P}_i=\sum_{m_i}\sum_{lurd=0}^{D-1}T^{[m_i]}_{lurd}{ \xi}^{m_i}\bar\eta_{{\rm L_i},l}^{p(l)}\bar\eta_{{\rm U_i},u}^{p(u)}\bar\eta_{{\rm R_i},r}^{p(r)}\bar\eta_{{\rm D_i},d}^{p(d)}.
\label{eq:Gprojector} 
\end{align}
where $\bar\eta_{{\rm L_i},l}$ is the $l-$th Grassmann variable on the leg $\rm L_i$ (similar for others). The fPEPS wave function in its Grassmann representation is written as
\begin{equation}
    \ket{\Psi}=\int_{\rm virtual} \prod_i {\bf P}_i\prod_{\langle i,j \rangle}{\bf G}_{ij}  \ket{0} , 
 \label{eq:fPEPS2}   
\end{equation}
where the subscript ``${\rm virtual}$'' means integrate over all the Grassmann variables on the virtual indices.

Assuming   ${\bf T}= \sum_{l_1\dots l_r}T_{l_1 \dots l_r} \eta_{l_1}^{p(l_1)}\cdots \eta_{l_r}^{p(l_r)}$ can be decomposed as $T_{l_1\dots l_r}=\sum_{uv}A_{l_1\cdots l_t, u}\delta_{uv}B_{v,l_{t+1}\cdots l_r}$.    Using $T_{l_1..l_r}=\sum_{uvxy}A_{l_1\cdots l_t, u}\delta_{ux}\delta_{xy}\delta_{yv}B_{v,l_{t+1}\cdots l_r}$, we have  ${\bf T}=\int_{\alpha_u \beta_v} {\bf ABG}$, where

\begin{equation}
 \begin{split}
&{\bf A}=\sum_{l_1..l_t,u}A_{l_1\cdots l_t, u}\eta_{l_1}^{p(l_1)}\cdots \eta_{l_t}^{p(l_t)}\bar \alpha^{p(u)}_u , \\
&{\bf B}=\sum_{l_{t+1}..l_r,v}B_{v,l_{t+1}\cdots l_r}\bar \beta^{p(v)}_v \eta_{l_{t+1}}^{p(l_{t+1})}\cdots\eta_{l_r}^{p(l_r)}, 
\label{eq:GA_3} \\
&{\bf G}=\sum_{xy}\delta_{xy} \beta^{p(y)}_y\alpha^{p(x)}_x=\sum_{x} \beta^{p(x)}_x\alpha^{p(x)}_x.
 \end{split}   
\end{equation}

Alternatively, we can absorb ${\bf G}$ into ${\bf A}$ or ${\bf B}$, to get other equivalent decomposition forms as given by Eq.~(\ref{eq:GA}), Eq.~(\ref{eq:GA_1}), or Eq.~(\ref{eq:GA_2}). This indicates that in the fPEPS expression Eq.~(\ref{eq:fPEPS2}), by absorbing ${\bf G}_{ij}$ into ${\bf P}_i$, we can reexpress fPEPS as
\begin{equation}
    \ket{\Psi}=\int_{\rm virtual} \prod_i {\bf T}_i \ket{0} , 
 \label{eq:fPEPS3}   
\end{equation}
where ${\bf T}_i$ has the same rank as ${\bf P}_i$.

\begin{figure}
    \centering
    \includegraphics[width=1\linewidth]{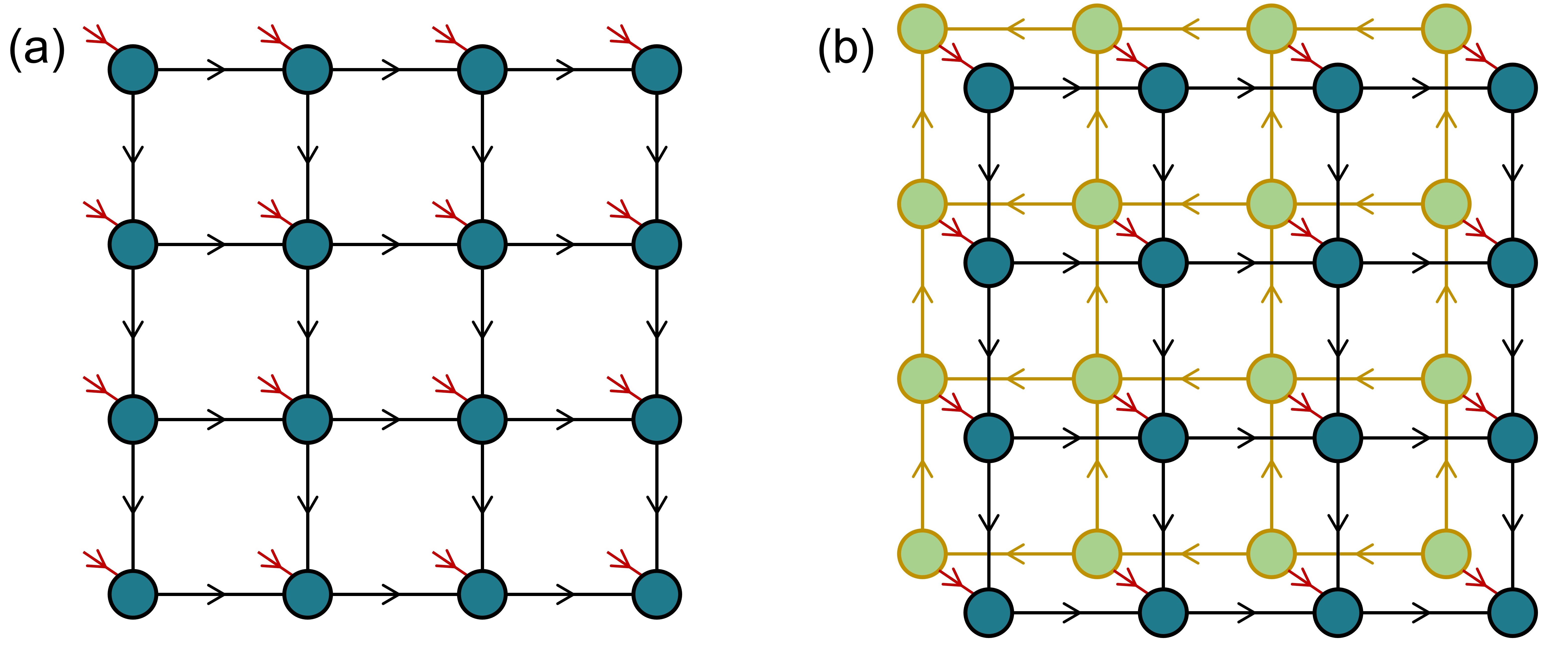}
    \caption{(a) Grassmann tensor network representation of fPEPS $| \Psi \rangle$. (b) Double-layer grassmann tensor network representation of the fPEPS norm (squared) $\langle \Psi | \Psi \rangle$. Arrow directions indicate contraction orders between $\eta$ and its conjugate $\bar \eta$.}
    \label{fig:fPEPSnorm}
\end{figure}

In Fig.~\ref{fig:fPEPSnorm}(a), we show the Grassmann tensor network representation for fPEPS. The arrow directions indicate the contraction orders to sum over virtual fermions, which can be chosen arbitrarily at the very beginning. Its norm can be computed by integrating out the physical degrees of freedom:
\begin{equation}
    \langle \Psi | \Psi \rangle=\int_{\rm virtual} \prod_i \Big(\int_{\rm physical}{{\bf T}^\dagger_i {\bf T}_i}\Big) \ket{0}, 
 \label{eq:fPEPSnorm}   
\end{equation}
where ${\bf T^{\dagger}_i}$ is the Hermitian conjugate, obtained from Eq.~(\ref{eq:conjugate_a}). The Grassmann tensor network for the norm is shown in  Fig.~\ref{fig:fPEPSnorm}(b), and it corresponds to a unique value.

\subsection{D. Operators in Grassmann form}

Now we consider the spinful fermionic operators  $c^{\dagger}_{\sigma}$ and $c_{\sigma}$ for the Hubbard model as a specific example of using the Grassmann form. The local basis elements for one site respectively are $\ket{0}$, $\ket{\uparrow}$, $\ket{\downarrow}$ and $\ket{\uparrow\downarrow}$, and

\begin{equation}
    \begin{split}
    & c^{\dagger}_{\uparrow}=\ket{\uparrow}\bra{0}+\ket{\uparrow\downarrow}\bra{\downarrow}, \\
   & c_{\uparrow}=\ket{0}\bra{\uparrow}+\ket{\downarrow}\bra{\uparrow\downarrow}, \\
   & c^{\dagger}_{\downarrow}=\ket{\downarrow}\bra{0}-\ket{\uparrow\downarrow}\bra{\uparrow}, \\
   & c_{\downarrow}=\ket{0}\bra{\downarrow}-\ket{\uparrow}\bra{\uparrow\downarrow}. 
    \end{split}
\end{equation}

Now we use four Grassmann variables to denote the four bases, $\alpha_{s}$ ($s=0,1,2,3$) for the corresponding ket $\ket{0}$, $\ket{\uparrow}$, $\ket{\downarrow}$ and $\ket{\uparrow\downarrow}$, and $\bar\xi_{t}$  ($t=0,1,2,3$) for the corresponding bra $\bra{0}$, $\bra{\uparrow}$, $\bra{\downarrow}$ and $\bra{\uparrow\downarrow}$.
Thus we can express spinful fermionic operators as  Grassmann tensors with  dimension $4\times 4$. Taking $c^{\dagger}_{\uparrow}$ as an example, $c^{\dagger}_{\uparrow}\doteq \sum_{st}A_{st}\alpha^{p(s)}_s \bar\xi_t^{p(t)}$, here $A_{st}=0$ except for $(s,t)=(1,0)$ which has $A_{10}=1$, $p(s)=1$ and $p(t)=0$, and for $(s,t)=(3,2)$ which has $A_{32}=1$, $p(s)=0$ and $p(t)=1$. Here $p(s)$ and $p(t)$ count the parity of the electron number in the bases.

Furthermore, given $c_{\uparrow}\doteq \sum_{mn}B_{mn}\alpha^{p(m)}_m \bar\xi_n^{p(n)}$, the hopping term $c^{\dagger}_{i,\uparrow}c_{j,\uparrow}$ is expressed as:
\begin{align}
   & c^{\dagger}_{i,\uparrow}c_{j,\uparrow}\doteq\sum_{stmn}A_{st}B_{mn}\alpha^{p(s)}_{i,s} \bar\xi_{i,t}^{p(t)}\alpha^{p(m)}_{j,m} \bar\xi_{j,n}^{p(n)}
    \nonumber \\
   &=\sum_{stmn}(-1)^{p(t)*p(m)}A_{st}B_{mn}\alpha^{p(s)}_{i,s}\alpha^{p(m)}_{j,m} \bar\xi_{i,t}^{p(t)} \bar\xi_{j,n}^{p(n)},
   \label{eq:hopping}
\end{align}
where $i$ and $j$ denote the Grassmann variables on site $i$ and $j$, respectively. Similarly we can express other operators including time evolution operators in the Grassmann form, and then we just need do Grassmann tensor operations to carry out fermionic tensor network computations.

\section{II. fPEPS algorithms}

Now we consider fPEPS computations on the Hubbard model. To fix the total electron number at  $N_{e}$, we impose  $U(1)$ symmetry on the local tensors ~\cite{singh2011tensor,bauer2011}. Below we consider two optimization schemes: the imaginary time evolution method, and the variational Monte Carlo method. The imaginary time evolution is conducted by the very efficient simple update method~\cite{jiang2008}, to find the approximate ground state. The variational Monte Carlo scheme provides a different approach for optimization and computing physical quantities, with initial states from the simple update method. 

\subsection{A. simple update imaginary time evolution}

\begin{figure}
    \centering
    \includegraphics[width=1\linewidth]{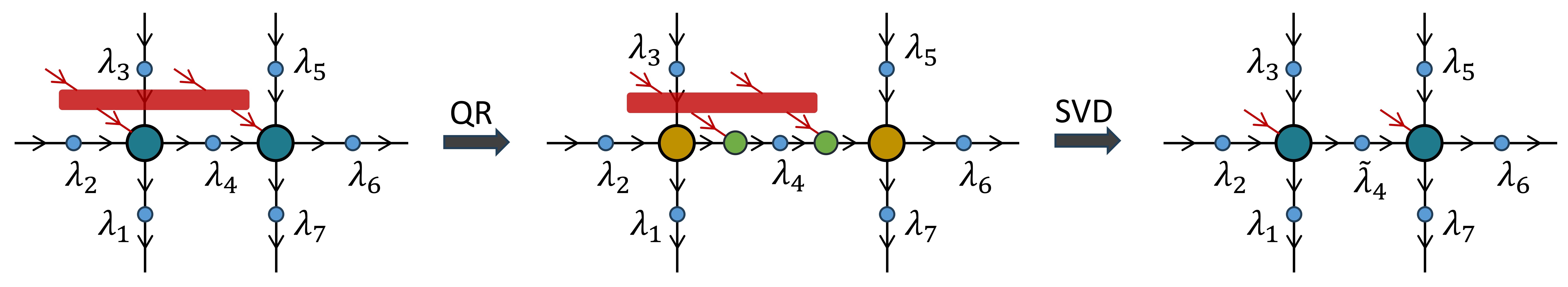}
    \caption{Simple update imaginary time evolution operator for fPEPS. Environment tensors are approximated by diagonal matrices $\lambda_i$ (blue circles), and are absorbed into tensors to be updated. Then the QR/LQ decomposition is done on the tensor pair, followed by a SVD on the subtensors (green circles)~\cite{wang2011}. Please see Ref.~\cite{liu2021} for a similar description with more details for spin PEPS. }
    \label{fig:fPEPSsvd}
\end{figure}

According to the imaginary time evolution method, the ground state is found by 
 \begin{equation}
     \ket{\Psi_0}= \lim_{\tau\rightarrow\infty} {\rm e}^{-\tau H}=\lim_{N\rightarrow\infty} {\rm e}^{-N{\rm d\tau} H},
 \end{equation}
 where $t=N {\rm d\tau}$.  The Hubbard Hamiltonian contains a set of two-body interaction terms, i.e., $H=\sum_{\langle ij \rangle}h_{ij}$, and then the evolution operator can be expressed with the Suzuki-Trotter expansion:
\begin{align}
    {\rm e}^{-{\rm d\tau} H}=\prod_{\langle ij \rangle} {\rm e}^{-{\rm d\tau} h_{ij}}+O(\tau^2).
\end{align}
The simple update method provides a very efficient approach for performing the imaginary time evolution~\cite{jiang2008}. In this way, the two-site evolution operator ${\rm e}^{-{\rm d\tau} h_{ij}}$ only acts on the site tensor $i$ and $j$, and the environmental effects from other sites are approximated by a series of diagonal matrices $\lambda_i$ on the dangling bonds of the tensors.  Note from Eq.~(\ref{eq:hopping}), the evolution operator has the following form:
\begin{align}
 {\rm e}^{-{\rm d\tau}h_{ij}}= \sum_{stmn}E_{smtn}\alpha^{p(s)}_{i,s}\alpha^{p(m)}_{j,m} \bar\xi_{i,t}^{p(t)} \bar\xi_{j,n}^{p(n)}.
\end{align}
Then the evolution operator acting on the fPEPS tensors is straightforward, which is graphically shown in Fig.~{\ref{fig:fPEPSsvd}}.  The SU is similar to that of bosonic/spin PEPS. It has a computational cost scaling as $O(D^5)$ by using QR/LQ decomposition  for  nearest neighbor interaction terms on the square lattice~\cite{wang2011}, as well as for next-nearest neighbor interactions~\cite{liu2021}, and thus allows us to reach a quite large bond dimensions $D$.

When using the SU method to update tensors, given the bond dimension $D$, we usually use an imaginary time step ${\rm d\tau}=0.01$ until the fPEPS is converged, that is the environment tensors $\lambda_i$ satisfy  $\frac{1}{P}\sum_{i=1}^{P}\frac{||\lambda_i{(\tau+{\rm d\tau})}-\lambda_i({\rm d\tau})||}{\lambda_i({\rm \tau})}<10^{-10}$. 
For spin models, when performing simple update, the large-$D$ PEPS state is usually initialized from a converged PEPS with small $D$, while for fermionic models we find  this tends to get trapped into a local minimum on $6\times 16$ and $8\times 16$ lattices. 
To escape this local minimum, we add pinning fields to target specified states. 

For the U(1)-charge symmetric fPEPS we consider here, we impose  magnetic field terms $-h_i {S}^z_i$ ($|h_i|\sim0.5$ and ${S}^z_i$ is the spin-$z$ component operator) or charge field terms $-\mu_i {n}_i$ ($\mu_i\sim0.5$ and ${n}_i$ is the particle number operator) in the Hamiltonian for the simple update. The pinning fields are imposed on  sites chosen to get a desired  magnetic pattern or charge pattern~\cite{zheng2017}. Taking the $8\times 16$ lattice as an example,  typically, we start from a random $D=8$ state and perform simple update under pinning fields and then increase $D$ directly from 8 to $16$, and further to $28$. For each   given $D=8$, 16 and 28, the simple update evolution is converged (i.e. $\lambda_i$ is converged as mentioned above). Next we remove the pinning field and continue the simple update with $D=28$ until convergence again. This fPEPS $D=28$ state is regarded as the stable state. With this $D=28$ state, we can get smaller-$D$ fPEPS state in a reverse process, i.e., by gradually decreasing $D$ from $28$ to smaller values through simple update and simultaneously ensuring $\lambda_i$ convergence for each $D$.

\subsection{B. Variational Monte Carlo}

In variational Monte Carlo, the energy function is expressed as
\begin{equation}
E=\frac{\ob{\Psi}{H}}{\norm{\Psi}}=\frac{1}{Z}\sum_{{\bf k}}|\Psi({\bf k})|^2 E_{\rm loc}({\bf k}) \doteq \langle E_{\rm loc}({\bf k}) \rangle~,
\label{eq:energy}
\end{equation}
where $\Psi({\bf k})$ is the amplitude of configuration $\ket{{\bf k}}$=$\ket{k_1k_2\cdots k_N}$, and 
\begin{equation}
E_{\rm loc}({\bf k})=\frac{\ele{{\bf k}}H{\Psi}}{\langle {\bf k}|\Psi\rangle}=\sum_{{\bf k}^{\prime}}\frac{\Psi({\bf k}^{\prime})}{\Psi({\bf k})}\ele{{\bf k}}H{{\bf k}^{\prime}}.
\end{equation}
The amplitude  $\Psi({\bf k})$ is computed through the standard boundary-MPS contraction method~\cite{verstraete2008,liu2021}.  Note  $\langle \cdots \rangle$ in Eq.(\ref{eq:energy}) denotes the Monte Carlo (MC) average. Additionally, the energy gradients with respect to the tensor elements are
\begin{equation}
     g_p=2\langle E_{\rm local}({\bf k}) O_p({\bf k})\rangle-2\langle E_{\rm local}({\bf k})\rangle \langle O_p({\bf k}) \rangle
     \label{eq:gradient}
\end{equation}
where $O_p({\bf k})=\frac{1}{\Psi({\bf k})}\Delta^{\dagger}_p({\bf k})$, and $\Delta_p({\bf k})$ is obtained by contracting a single-layer tensor network with the configuration $\ket{{\bf k}}$ except for the element at the corresponding position $p$~\cite{liu2021}, and $\Delta^{\dagger}_p({\bf k})$ is its Hermitian conjugate defined as in Eq.~(\ref{eq:conjugate_a}).

\subsubsection{1. Monte Carlo sampling procedure}

In the MC sampling, we sample the configurations in the subspace $N_{\uparrow}=N_{\downarrow}=\frac{1}{2}N_{e}$, where $N_{\uparrow}$ ($N_{\downarrow}$) is the number of spin-up (down) particles. This is equivalent to enforcing $U(1)\times U(1)$ symmetry on the wave function. The configurations are generated using Metropolis' algorithm. 
Specifically, assuming the current configuration is $\ket{{\bf k}_a}=\ket{k_1 \cdots k_ik_{i+1}\cdots k_{N}}$  satisfying $N_{\uparrow}=N_{\downarrow}=\frac{1}{2}N_{f}$, we propose a trial configuration  $\ket{{\bf k}_b}=\ket{k_1\cdots k'_ik'_{i+1}\cdots k_{N}}$ obtained by flipping the states on a nearest-neighbor pair of site $i$ and $i+1$, to be accepted according to Metropolis' algorithm. To generate configurations quickly, we loop over the rows and attempt to flip each pair according to the Metropolis acceptance criterion~\cite{liu2021}: 
\begin{equation}
    P={\rm min}\Bigg[1, \frac{|\Psi({\bf k}_b)|^2}{|\Psi({\bf k}_a)|^2}\Bigg].
\end{equation}
The trial configuration $\ket{{\bf k}_b}$ will be accepted as a new configuration if a randomly chosen number from the uniform distribution in the interval [0,1) is smaller than the probability $P$. Otherwise,  the trial configuration $\ket{{\bf k}_{b}}$ is rejected, and another trial configuration $\ket{{\bf k}_{b}^{\prime}}$ is considered.  To keep $N_{\uparrow}=N_{\downarrow}=\frac{1}{2}N_{e}$, the trial states are constrained. For example, if the current states on sites $i$ and $i+1$ are $\ket{\uparrow_{i},\downarrow_{i+1}}$, then the trial states $\ket{k'_ik'_{i+1}}$ can only be $\ket{\uparrow_{i}\downarrow_{i},0_{i+1}}$, $\ket{0_{i},\uparrow_{i+1}\downarrow_{i+1}}$, or $\ket{\downarrow_{i},\uparrow_{i+1}}$.  A Monte Carlo sweep is defined as a sweep over all horizontal  and  vertical nearest-neighbor pairs through the lattice, and the physical quantities are measured after each Monte Carlo sweep.

\subsubsection{2. Stochastic optimization}

The energy gradients can be evaluated via Monte Carlo sampling, and thus it is natural to optimize the fPEPS by gradient methods. We consider two gradient-based methods: stochastic gradient descent (SGD)~\cite{sandvik2007,liu2017,liu2021} and stochastic reconfiguration (SR)~\cite{SR1998,Neuscamman2012,directSam2021}. 

 \begin{figure}[htbp]
 \centering
\includegraphics[width=\columnwidth]{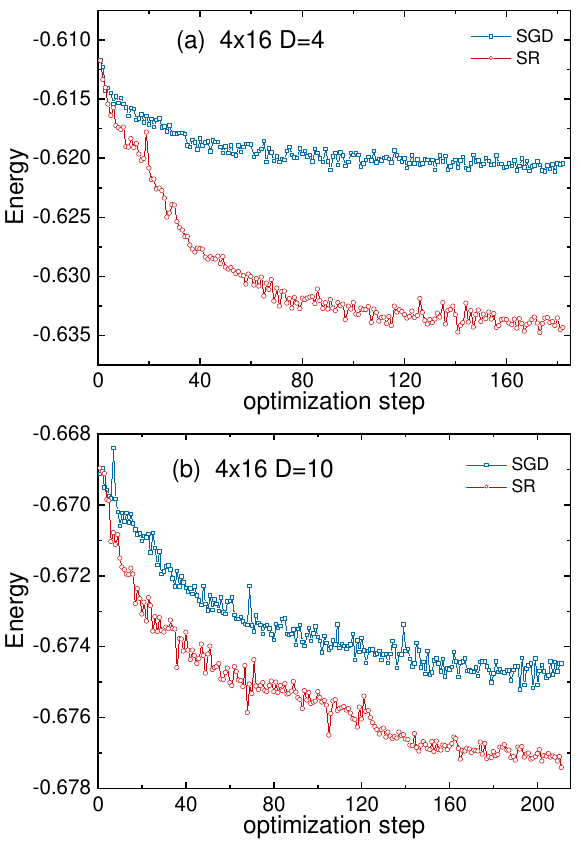}
 \caption{Gradient-based optimization of energy with SGD and SR methods, for the $4\times 16$ $U=8$ Hubbard model with $N_e=56$ electrons, using fPEPS (a) $D=4$ and (b) $D=10$. }
 \label{fig:SR_SGD}
 \end{figure}

For SGD, tensor elements are updated following the (negative) sign of their gradients:
\begin{equation}
  x_p(\tau +1)= x_p(\tau )-r_p \cdot {\rm d}\tau\cdot {\rm{sgn}}(g_p),
\end{equation}
where $\tau$ is the number of optimization steps, and $r_p$ is an independent random number in the interval $(0,1)$ for each tensor element $x_p$.  The parameter ${\rm d}\tau$ is the step size, setting the variation range for an element. Starting from the simple update state, the step length ${\rm d}\tau$ can be gradually tuned from 0.005 to a smaller one like 0.001~\cite{liu2021}. 

For SR, it optimizes the tensors on the fPEPS manifold, which is equivalent to the imaginary time evolution using the time-dependent variational principle~\cite{SR1998}. In this method,  we need to solve a system of linear equations: 
\begin{equation}
\sum_{q}S_{pq}\dot{x}_q=g_p,   
\label{eq:sr}
\end{equation}
where $S_{pq}=\langle O_p({\bf k}) O_q({\bf k})\rangle-\langle O_p({\bf k})\rangle \langle O_q({\bf k})\rangle$. 
To avoid explicitly constructing the ${\bf S}$ matrix, we solve the above equation in an iterative way~\cite{Neuscamman2012}, and  we only need to know how to map a vector ${\bf \dot{x}}$ to ${\bf y}$ under the action of ${\bf S}$ to realize  ${\bf y=S\dot{x}}$. Specifically,
\begin{equation}
    y_p=\sum_{q}S_{pq}\dot{x}_q=y_p^{(1)}-y_p^{(2)},
    \nonumber
\end{equation}
where
\begin{align}
       & y_p^{(1)}=\frac{1}{N_{\rm MC}}\sum_{\bf k} O_p ({\bf k}) \Big[\sum_q O_q({\bf k})\dot{x}_q\Big],     \nonumber\\
       &y_p^{(2)}=\langle O_p({\bf k})\rangle \sum_{q}\langle O_q({\bf k}) \rangle \dot{x}_q,     \nonumber
\end{align}
 where $\langle O_p({\bf k})\rangle\equiv\frac{1}{N_{\rm MC}}\sum_{\bf k}O_p({\bf k})$ is the Monte Carlo average of  $O_p({\bf k})$, and $N_{\rm MC}$ is the number of Monte Carlo sweeps.  Note $O_p({\bf k})$ can be stored trivially on different CPU processors, while still being convenient to iteratively solve  Eq.~(\ref{eq:sr}) (a diagonal shift $0.001$ for regularizing  the ${\bf S}$ matrix  is also added). Once we obtain the solution $\dot{x}_p$, the tensor elements are updated as
\begin{equation}
  x_p(\tau+1)= x_p(\tau)-  {\rm d}\tau \cdot\dot{x}_p ,
\end{equation}
where ${\rm d}\tau$ is the step size, which can be tuned from ${\rm d}\tau=0.05$ to 0.01.
 
In Fig.~\ref{fig:SR_SGD}, we show the energy variation with respect to the optimization step by SGD and SR methods for the $4\times 16$ lattice with Hubbard interaction $U=8$ at hole doping $n_h=0.125$, using fPEPS $D=4$ and $D=10$. For all cases, we start from the states from the simple update method, and the number of Monte Carlo sweeps for each optimization step is about 20000. The step size is adjusted from $0.005$ to $0.001$ for SGD and from $0.05$ to $0.01$ for SR according to the energy behavior, ensuring good optimization performance. Clearly, for both $D=4$ and $D=10$, the SR method works better than the SGD method. Therefore, all of the results presented from gradient-based optimization (GO) use the SR optimization.

\subsubsection{3. Accuracy comparison of gradient optimization and simple update}
\begin{figure}[tbp]
 \centering
 \includegraphics[width=0.5\textwidth]{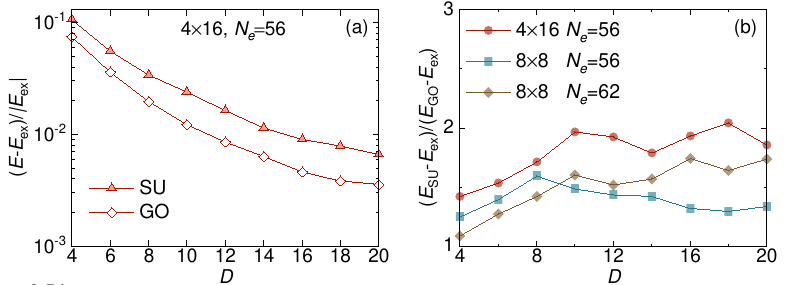}
 \caption{ (a) The relative energy errors of SU and GO results with respect to $D$ for the $4\times 16$ Hubbard model at $U=8$ with electron numbers $N_e=56$. The extrapolated DMRG energy is used for the reference. (b) The ratio between the SU energy error and the GO energy error for each bond dimension $D$ for various systems. }
 \label{fig:energy4x16}
 \end{figure}

\begin{figure}[tbp]
 \centering
 \includegraphics[width=0.99\columnwidth]{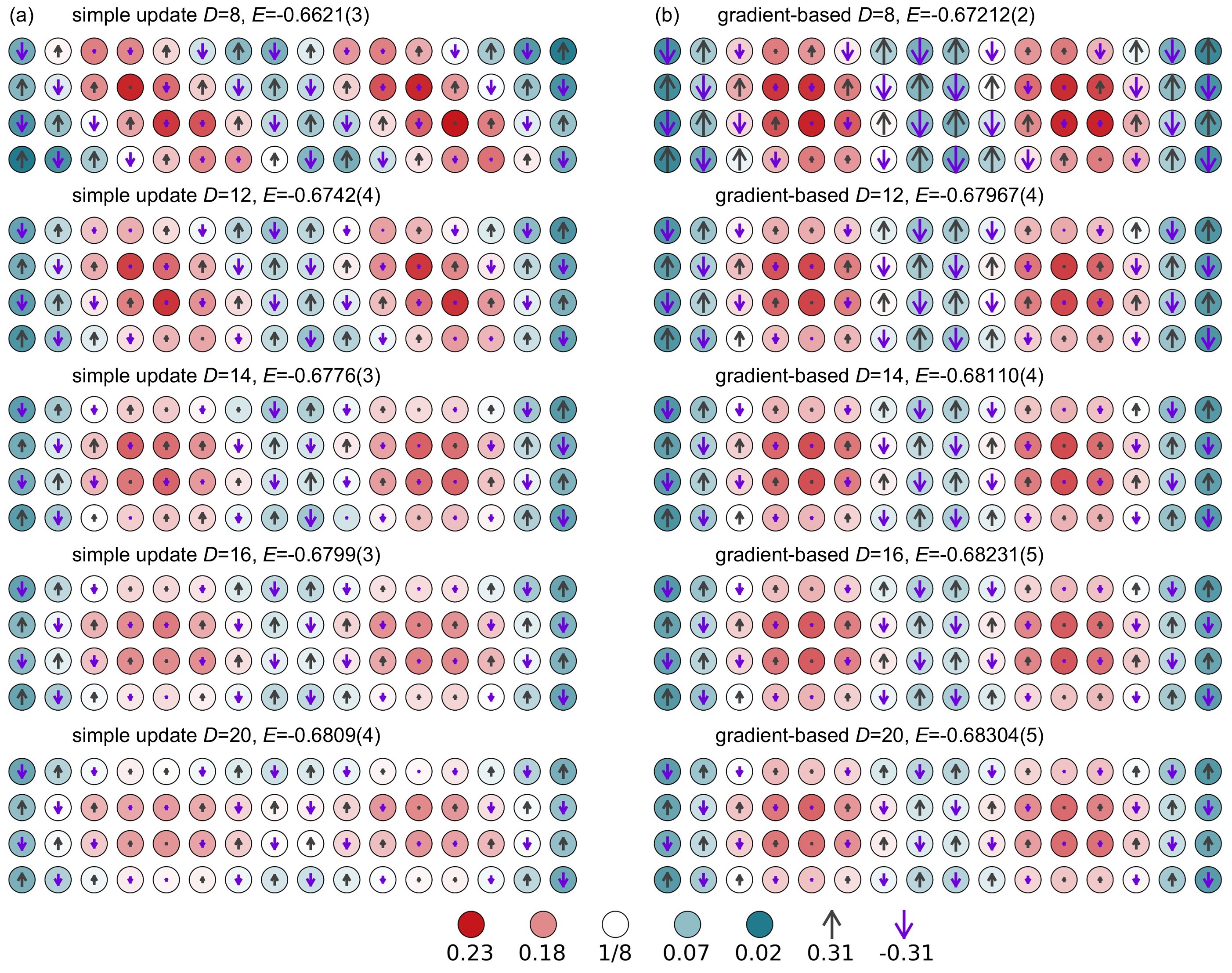}
 \caption{ The charge and spin pattern on $4\times 16$ with $U=8$ and $N_e=56$. The left part shows the results obtained from the simple update, and the right is from gradient-based optimization. For the same $D$, gradient-based optimization is initialized by the simple update fPEPS state.}
 \label{fig:pattern4x16}
 \end{figure}

 \begin{figure}[tbp]
 \centering
 \includegraphics[width=0.99\columnwidth]{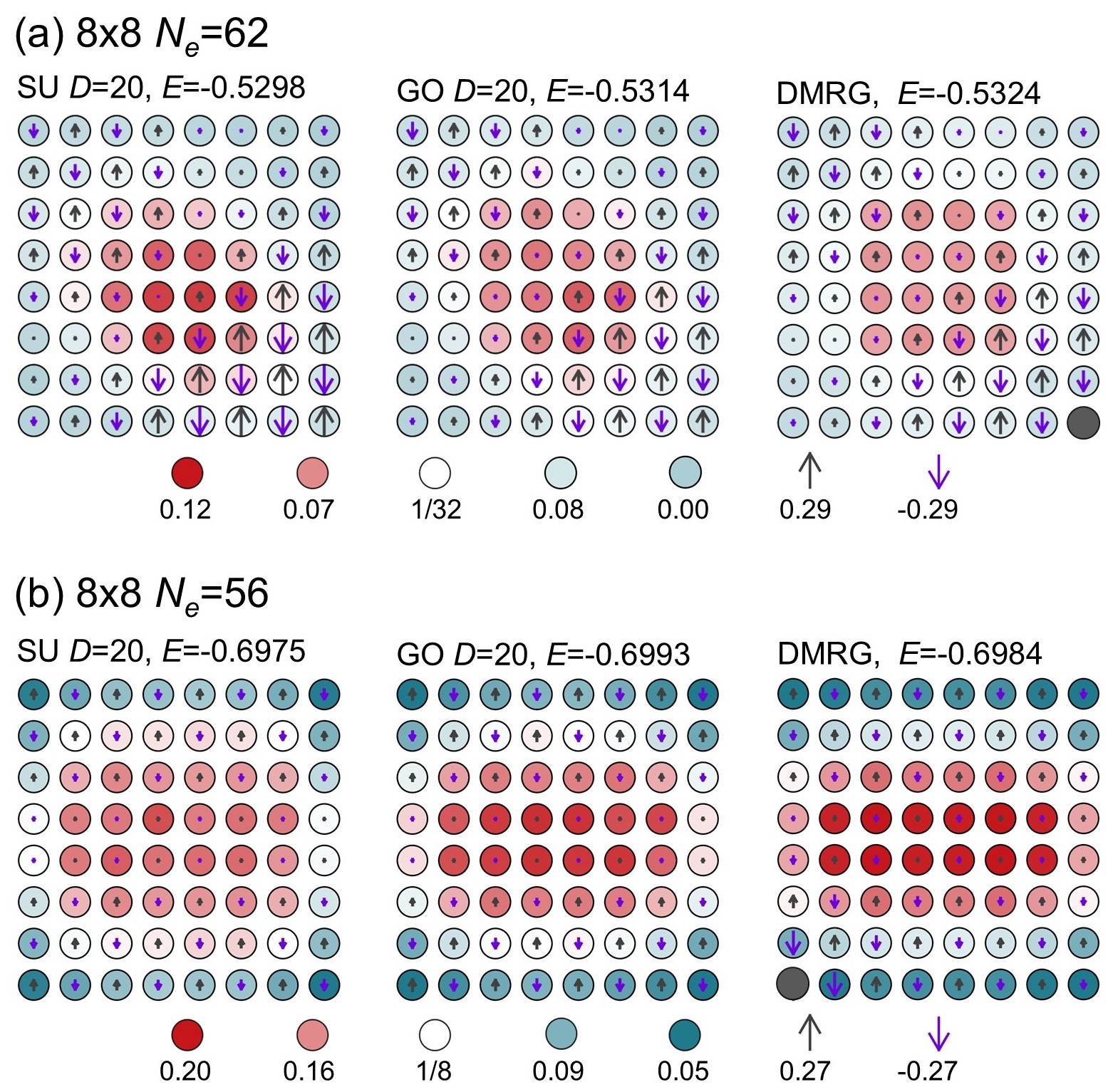}
 \caption{ The charge and spin pattern on the $8\times 8$ lattice with $U=8$ and (a) $N_e=62$ and (b) $N_e=56$. From left to right, the results are respectively obtained from the simple update (SU), gradient-based optimization (GO), and SU(2)-DMRG with $m=32000$. For DMRG results, local spin $z$-component moments are zero due to SU(2) symmetry, and the presented arrows denote the spin correlation values $|\frac{1}{3}\langle {\bf S}_{\rm ref}\cdot {\bf S}_{i,j}\rangle|^{1/2}$ using the reference site (filled black circles), and each arrow direction (up or down) depends on the sign of $\langle {\bf S}_{\rm ref}\cdot {\bf S}_{i,j}\rangle$. }
 \label{fig:pattern8x8}
 \end{figure}

Now we compare the accuracy of the energy obtained from gradient optimization and simple update.  In Fig.~\ref{fig:energy4x16}(a), we show the energies from SU and GO methods. For SU, the energy relative error can be as small as 0.0055 with $D=20$. With SU states as starting points, GO can further improve the accuracy, for example, with the error going down to 0.0037 for $D=20$.  By comparing the energy errors between the GO and SU methods for various bond dimensions $D$ on different system sizes $L_y\times L_x$ including $4\times 16$ and $8\times 8$, we find that the energy error of SU is only twice as large as that of GO, as shown in Fig.~\ref{fig:energy4x16}(b). 
 
 Meanwhile, we observe that both SU and GO yield nearly identical patterns of physical quantities for large $D$, as shown in Fig.~\ref{fig:pattern4x16} for $4\times 16$ and  Fig.~\ref{fig:pattern8x8} for $8\times 8$, with only slight quantitative differences. These findings suggest SU with large-$D$ is able to converge essential aspects of physics.

\subsubsection{4. Monte Carlo convergence and contraction convergence}

\begin{figure}[htb]
\centering
\includegraphics[width=0.99\columnwidth]{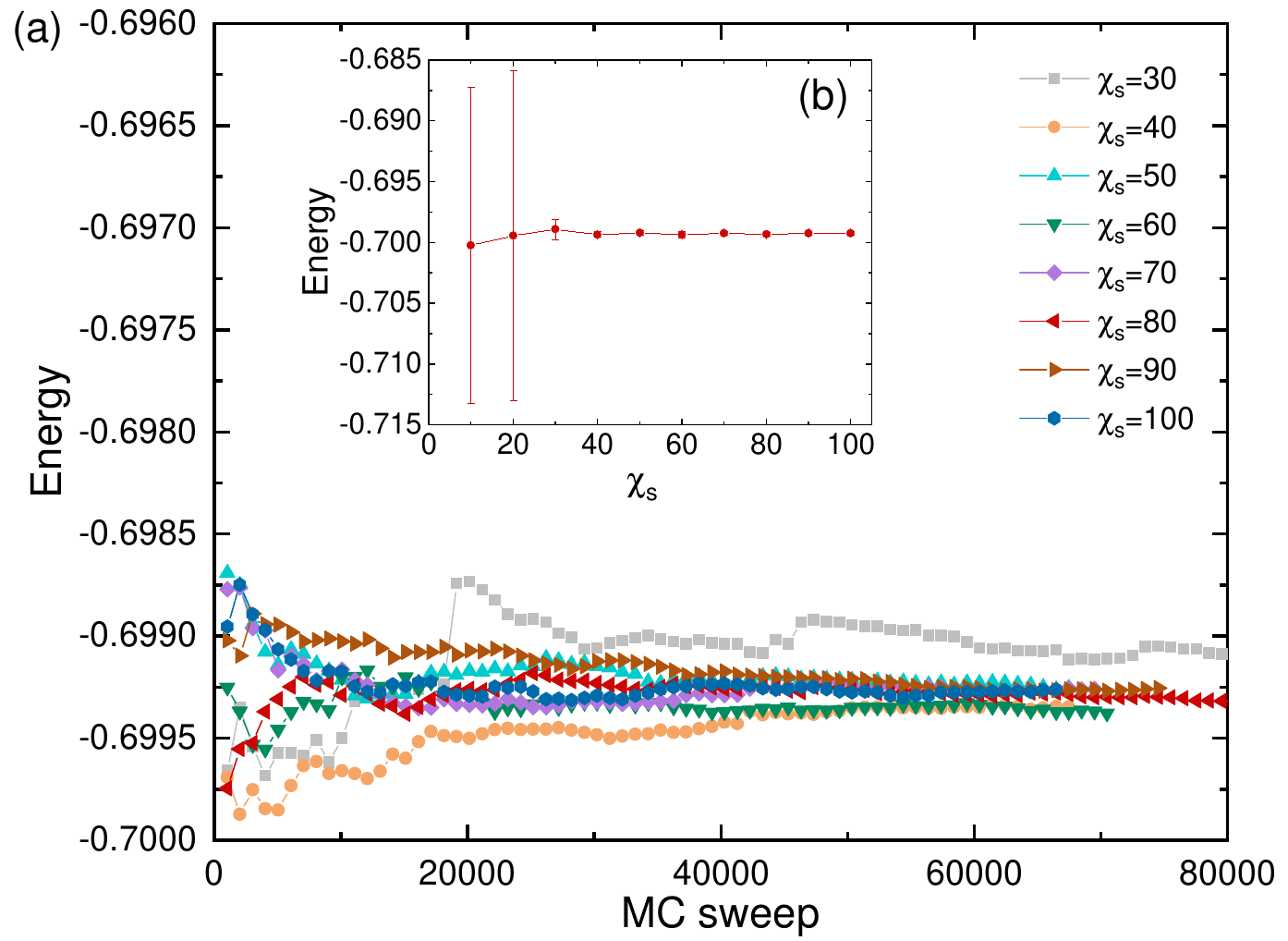}
\caption{For the  $8\times 8$ Hubbard model at $U=8$ and $N_e=56$ using fPEPS $D=20$, the energy convergence behavior with respect to (a) MC sampling for  different cutoff $\chi_s$, and (b) cutoff $\chi_s$ with maximal samples from MC, and error bars are mean standard errors which are evaluated from multiple bins from MC sampling, in which each bin contains 10000 samples.} 
\label{fig:8x8MCsweeps} 
\end{figure}

\begin{figure}[tbp]
 \centering
 \includegraphics[width=0.99\columnwidth]{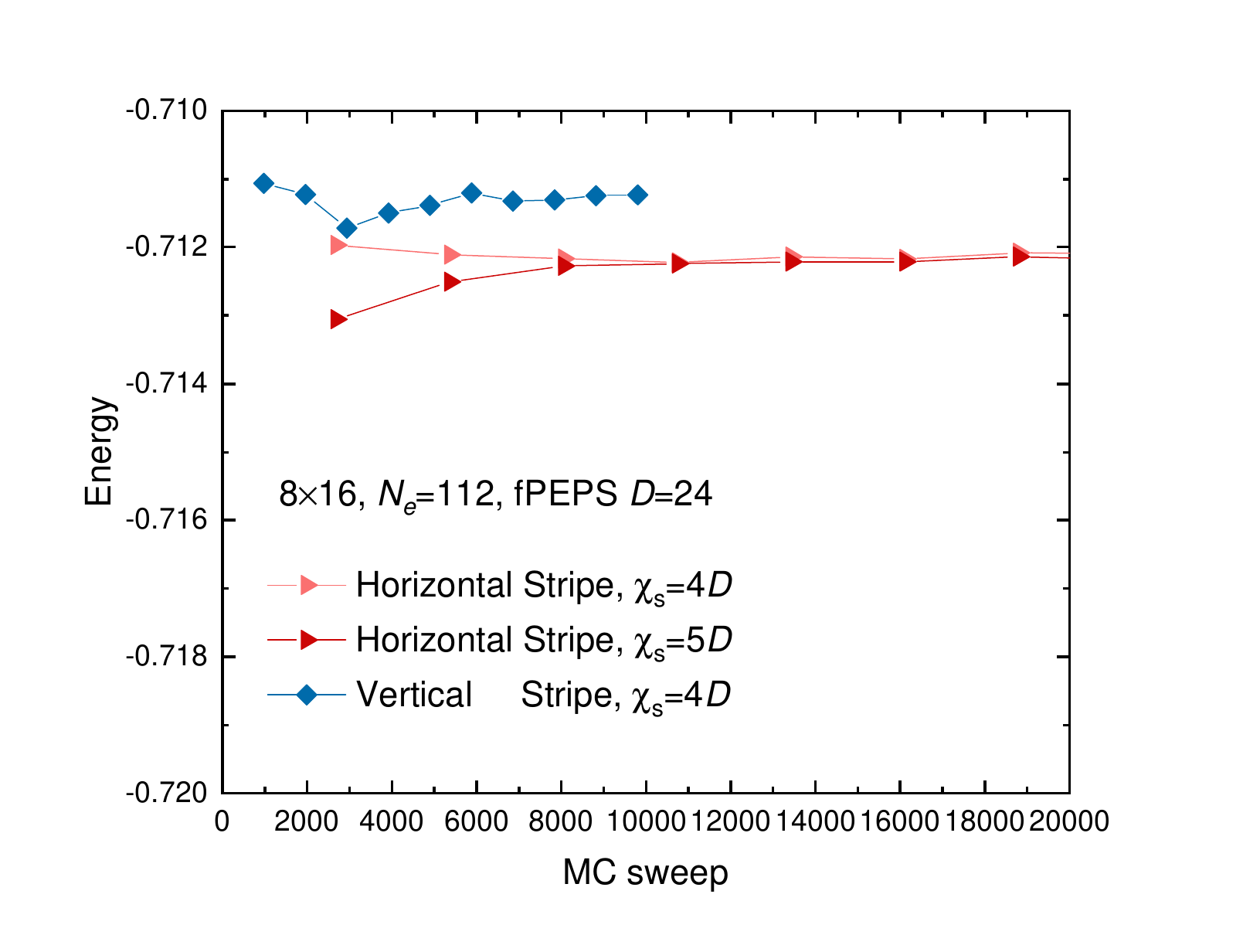}
 \caption{ Energy convergence with respect to MC sampling on the $8\times 16$ Hubbard model at $N_e=112$ and $U=8$, for the horizontal stripe state  with cutoff $\chi_s=96$ (light red) and $\chi_s=120$ (dark red), and vertical stripe state with $\chi_s=96$ (blue). }
 \label{fig:8x16MCsweeps}
 \end{figure}

 \begin{figure}[tbp]
 \centering
 \includegraphics[width=0.99\columnwidth]{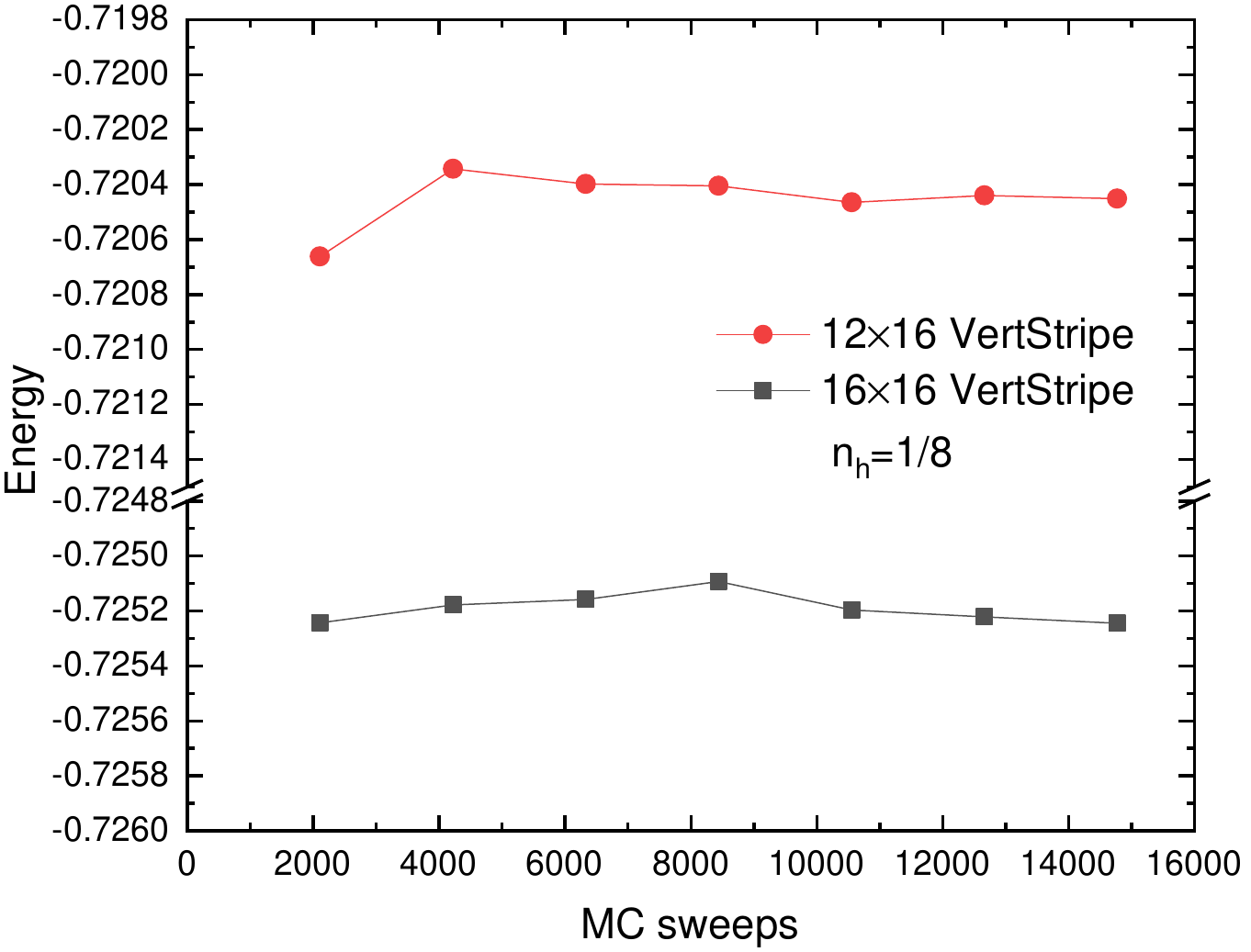}
 \caption{ Energy convergence with respect to MC sampling for $12\times 16$ and $16\times 16$ Hubbard model at $U=8$ with hole doping $n_h=1/8$ (that is $N_e=168$ and $N_e=224$ correspondingly) for vertical stripe states, using $D=18$ with cutoff $\chi_s=4D=72$. }
 \label{fig:16x16MCsweeps}
 \end{figure}

\begin{figure*}[tbp]
 \centering
 \includegraphics[width=1.99\columnwidth]{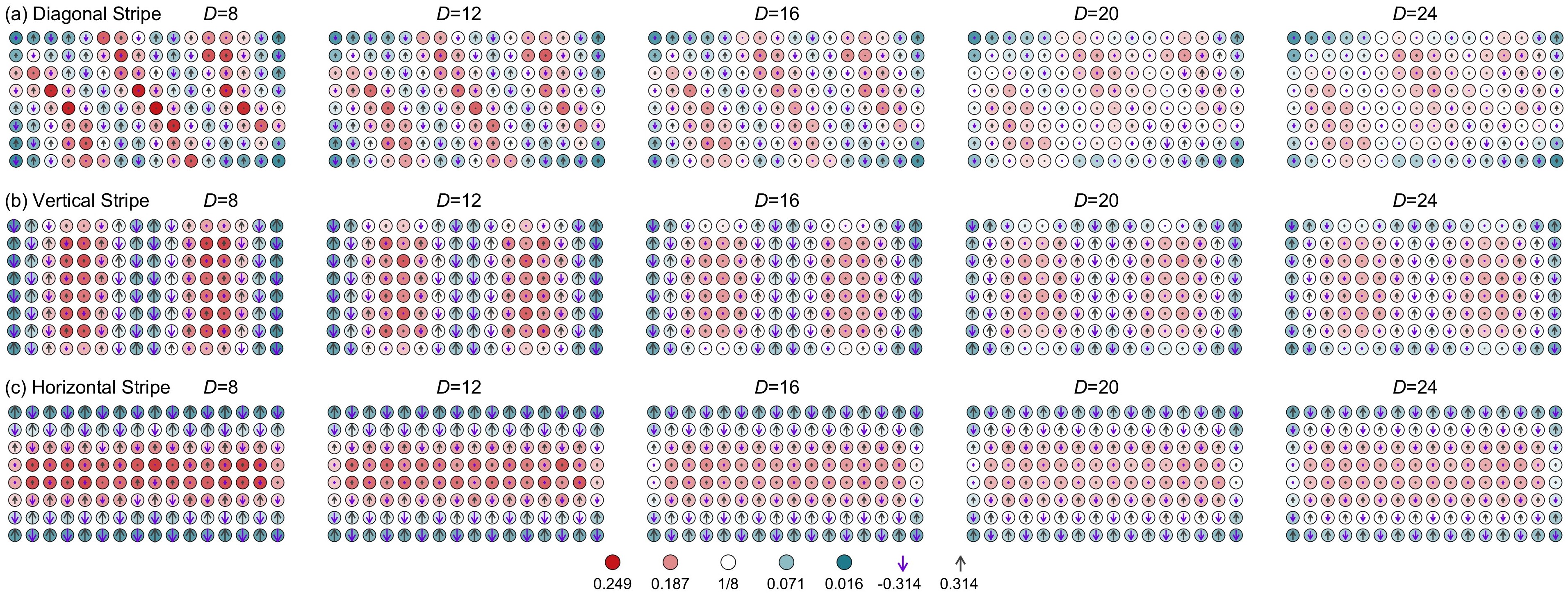}
 \caption{ The hole density and spin distribution on  the $8\times 16$ Hubbard model with $U=8$ and $N_e=112$, from fPEPS with different bond dimensions $D$. }
 \label{fig:pattern8x16}
 \end{figure*}

Now we consider the convergence behaviour of MC sampling, as well as of the cutoff $\chi_s$ used for contracting single-layer tensor networks (the amplitude of a configuration) by the boundary-MPS method.  
Here we take the $8\times 8$ Hubbard model with $U=8$  and $N_e=56$ as an example, using the fPEPS $D=20$ ground state. Fig.~\ref{fig:8x16MCsweeps}(a) and (b) show the energy convergence behavior with respect to MC sweeps and $\chi_s$, respectively.  For a given large $\chi_s$ presented in Fig.~\ref{fig:8x16MCsweeps}(a),  we can see the energy (persite) can be well evaluated by around $10000$ MC sweeps, with an statistical error smaller than $3\times 10^{-4}$. 

As another example we consider the  $8\times 16$ Hubbard model with $U=8$  and $N_e=112$, using fPEPS $D=24$. For the case $8\times 8$, it has been shown $\chi_s=3D=60$ is enough to converge the energy [Fig.~\ref{fig:8x8MCsweeps}(a)]. Here we consider larger $\chi_s$ with $\chi_s=4D=96$ and $\chi_s=5D=120$ for $8\times 16$, and it shows $\chi_s=4D$ is good for convergence. Additionally, $10000$ MC sweeps also work well to have a small statistical error about $3\times 10^{-4}$.  

We further extend the MC convergence analysis to larger sizes ($12 \times 16$ and $16\times 16$) as shown in Fig.~\ref{fig:16x16MCsweeps}, where similar MC sweep counts (around $10000$) are sufficient to maintain adequate accuracy.
Notably, as the wavefunction approaches the ground state, small sample sizes already produce precise energy estimates regardless of the system size. This phenomenon stems from the energy zero variance principle: For an exact ground state $|\Psi\rangle$ satisfying $H|\Psi\rangle=E_0|\Psi\rangle$, the local energy $E_{\rm loc}({\bf k})$ becomes configuration-independent, i.e. $E_{\rm loc}({\bf k})\equiv E_0$. Consequently, as the quantum state approaches the ground state, its energy estimator becomes increasingly immune to statistical sampling errors, enabling size-independent measurement precision (in the energy per site) even with limited samples.

\section{III. Pattern evolution with bond dimension $D$}

In Fig.~\ref{fig:pattern4x16}, we have shown how the pattern evolves with respect to the fPEPS bond dimension $D$. Here we consider the case of $8\times 16$ at hole doping $n_h=1/8$ at $U=8$. In Fig.~\ref{fig:pattern8x16}, we show the pattern evolution for the diagonal, vertical and horizontal stripes. In each case, small-$D$ results exhibit apparently ordered patterns. Increasing $D$, the orders are softened due to more quantum fluctuations being included in the fPEPS ansatz.  For larger sizes, the patterns show a similar evolution of features as a function of $D$.

  \begin{table*}[htp]
  \centering
\caption {fPEPS and DMRG energies persite for $4\times 4$ at half filling ($N_e=16$) and $n_h=1/8$ hole doping ($N_e=14$ ) at Hubbard repulsive interaction $U=8$. For DMRG results, the truncated weights for bond dimension $m=2500$ are $4.12\times 10^{-9}$ and $7.55\times 10^{-8}$ for half-filling and $n_h=1/8$, respectively. DMRG results with $m=\infty$ denote extrapolated energies using truncated weights. }
	\begin{tabular}{lllll|lllll}\hline\hline
 
  \multicolumn{5}{c|}{$4\times 4$, half-filling} & \multicolumn{5}{c}{$4\times 4$, $n_h=1/8$} \\
       \hline
    $D$&   fPEPS-SU    &   fPEPS-GO  &  $m$ & DMRG & $D$&   fPEPS-SU    &   fPEPS-GO  &  $m$ & DMRG \\
  
 4 & -0.4174(2) & -0.42115(3) & 500  &-0.42551504 & 4  & -0.5497(6) & -0.5892(2)  & 500   & -0.63241788\\
 6 & -0.4204(6) & -0.42416(7)  & 1000 &-0.42552519 & 6  & -0.5850(4) & -0.61225(7)   & 1000  & -0.63260460\\
 8 &-0.4238(5)  &-0.42508(4)  & 1500 &-0.42552583 & 8  &-0.6016(3) & -0.62314(6)     & 1500  & -0.63261660\\
10 &-0.4236(3) &-0.425429(8) & 2000 &-0.42552589 & 10 &-0.6104(4) & -0.62725(4)   & 2000   & -0.63261805\\
12 &-0.4241(3) &-0.425493(3) & 2500 &-0.42552590 & 12 &-0.6125(5) &-0.63159(3)    & 2500   &-0.63261830\\
14 &-0.4242(5) &-0.425508(5) & $\infty$ & -0.42552596(1) & 14 &-0.6154(4) &-0.63206(1) & $\infty$ & -0.6326190(1)\\
16 &-0.4244(2) & -0.425519(6) &      &            & 16 &-0.6152(4) &-0.63234(2) \\
          
             \hline\hline
	\end{tabular}
\label{tab:peps-4x4}
\end{table*}

  \begin{table*}[htp]
  \centering
\caption {fPEPS and DMRG energies persite for the Hubbard model at $U=8$. For $6\times 6$, truncated weights from $m=24000$ for $n_h=1/18$ ($N_e=34$) and $n_h=1/9$ ($N_e=32$) are $6.01\times 10^{-6}$ and $3.26\times 10^{-5}$, respectively. For  $4\times 16$, truncated weights from $m=12000$ is $1.01\times 10^{-5}$.  DMRG results with $m= \infty$ denote extrapolated energies from truncated weights, not using the results  from the presented maximal bond dimensions. }

\begin{tabular}{lllll|lllll|lllll }\hline\hline
 
  \multicolumn{5}{c|}{$6\times 6$, $n_h=1/18$} & \multicolumn{5}{c|}{$6\times 6$, $n_h=1/9$} & \multicolumn{5}{c}{$4\times 16$, $n_h=1/8$} \\
       \hline
    $D$&   fPEPS-SU    &   fPEPS-GO  &  $m$ & DMRG & $D$&   fPEPS-SU    &   fPEPS-GO  &  $m$ & DMRG & $D$&   fPEPS-SU    &   fPEPS-GO  &  $m$ & DMRG  \\
  
 4 & -0.5132(5) & -0.5273(3) & 16000  & -0.5598359    & 4  & -0.5786(4) & -0.6007(7)  & 16000    & -0.6570640  & 4 & -0.6120(8) & -0.63383(4) & 8000  &-0.6852680 \\
 6 & -0.5363(3) & -0.5448(1)  & 18000 & -0.5598593  & 6  & -0.6200(6) & -0.6309(4)    & 18000    & -0.6571482 & 6 & -0.6473(4) & -0.66062(6)  & 9000 &-0.6853034\\
 8 &-0.5438(5) & -0.5509(1)  & 20000  & -0.5598766   & 8  & -0.6312(5)  &-0.6415(2)   & 20000    & -0.6572133 & 8 & -0.6621(3) & -0.67212(2)  & 10000 &-0.6853307\\
10 &-0.5486(2) & -0.5542(2) & 22000   & -0.5598898      & 10 & -0.6388(5) &-0.6448(2) & 22000    & -0.6572647 & 10 & -0.6690(4) & -0.67712(5) & 11000 & -0.6853523 \\
12 &-0.5510(4) &-0.5561(2) & 24000   &  -0.5599001    & 12 & -0.6423(3) &-0.6485(1)   & 24000    & -0.6573063 &12 & -0.6742(4) & -0.67967(4) & 12000 & -0.6853698\\
14 &-0.5528(3) &-0.55766(7)  & $\infty$ & -0.559969(16)  & 14 & -0.6455(3) &-0.6513(2) & $\infty$& -0.657675(82) &14 & -0.6776(2) & -0.68110(4) &$\infty$ & -0.685544(38)\\
16 & -0.5531(4) &-0.55827(3) &      &     & 16 & -0.6470(4) & -0.65327(5)  & &            &16 & -0.6799(3) & -0.68231(5)  &      &    \\
18 &-0.5539(2) &-0.55880(8)  &      &     & 18 & -0.6475(3) & -0.65383(6) & &            &18 & -0.6803(3) & -0.68272(2)  &      &  \\
20 &-0.5542(2) &-0.55908(7) &      &     & 20 & -0.6477(2) & -0.65475(8) & &            &20 & -0.6809(4) & -0.68304(5)  &      &  \\
             \hline\hline
	\end{tabular}
\label{tab:peps-6x6}
\end{table*}

\bibliography{fPEPS-hubbard-main}

\end{document}